\documentclass[12pt]{article}
\usepackage{amsmath,amssymb}
\usepackage{graphicx}
\newcommand{\eqdef}{\stackrel{\text{def}}{=}}

\newcommand{\n}{\nonumber \\}
\newcommand{\bm}{\boldsymbol}
\newcommand{\ignore}[1]{}
\newcommand{\lbr}{[\![}
\newcommand{\rbr}{]\!]}
\newcommand{\lpar}{(\!(}
\newcommand{\rpar}{)\!)}
\newcommand{\td}{\tilde{d}}
\numberwithin{equation}{section}
\newcommand{\Romannumeral}[1]{\uppercase\expandafter{\romannumeral#1}}

\newcommand{\norm}[1]{|\!|#1|\!|}

\newtheorem{theo}{\bf Theorem}[section]

\newtheorem{rema}[theo]{\bf Remark}

\newtheorem{prop}[theo]{\bf Proposition}
\newtheorem{defi}[theo]{\bf Definition}

\allowdisplaybreaks[4]

\begin{document}

\baselineskip=20pt
\newcommand{\preprint}{
\vspace*{-20mm}\begin{flushleft}\end{flushleft}
}
\newcommand{\Title}[1]{{\baselineskip=26pt
  \begin{center} \Large \bf #1 \\ \ \\ \end{center}}}
\newcommand{\Author}{\begin{center}
  \large \bf 
  Ryu Sasaki${}$ \end{center}}
\newcommand{\Address}{\begin{center}
     Department of Physics and Astronomy, Tokyo University of Science,
     Noda 278-8510, Japan
        \end{center}}
\newcommand{\Accepted}[1]{\begin{center}
  {\large \sf #1}\\ \vspace{1mm}{\small \sf Accepted for Publication}
  \end{center}}

\preprint
\thispagestyle{empty}

\Title{Towards verifications of Krylov complexity}

\Author

\Address
\vspace{1cm}

\begin{abstract}
Krylov complexity is considered to provide a measure of the growth of operators evolving under 
Hamiltonian dynamics.
The main strategy is the analysis of the structure of Krylov subspace $\mathcal{K}_M(\mathcal{H},\eta)$
spanned by the multiple
applications of the Liouville operator  $\mathcal{L}$ defined by the commutator in terms of a Hamiltonian
$\mathcal{H}$, $\mathcal{L}:=[\mathcal{H},\cdot]$ acting on an operator $\eta$,
$\mathcal{K}_M(\mathcal{H},\eta)=\text{span}\{\eta,\mathcal{L}\eta,\ldots,\mathcal{L}^{M-1}\eta\}$.
For a given inner product $(\cdot,\cdot)$ of the operators, 
the  orthonormal basis $\{\mathcal{O}_n\}$ is constructed
from $\mathcal{O}_0=\eta/\sqrt{(\eta,\eta)}$ by Lanczos algorithm.
The moments $\mu_m=(\mathcal{O}_0,\mathcal{L}^m\mathcal{O}_0)$ are closely related to the important 
data $\{b_n\}$ called Lanczos coefficients.
I present the exact and explicit expressions of the moments $\{\mu_m\}$ for 16 quantum mechanical 
systems which are {\em exactly solvable both in the Schr\"odinger and Heisenberg pictures}.
The operator $\eta$ is  the variable of the eigenpolynomials. 
Among them six systems 
show a clear sign of `non-complexity' as vanishing higher Lanczos coefficients $b_m=0$, $m\ge3$.
\end{abstract}

\begin{center}
Published:\quad Progress of Theoretical and Experimental Physics, \\
Volume 2024, Issue 6, June 2024, 063A01, \\
Doi: https://doi.org/10.1093/ptep/ptae073
\end{center}

%
%
\section{Introduction}
\label{sec:intro}
In the study of quantum chaos, Krylov complexity is proposed as a measure of the growth of operators in
the Heisenberg picture \cite{parker, barbon, dymarsky, rabino, caputa, yyang,pratik}.
The definition of Krylov complexity 
is universal, that is, applicable to a very complicated system as 
well as an extremely ordered system, for example, an exactly solvable system in the Heisenberg picture.
As materials to support the validity of the concept of Krylov complexity {\em by contrast} 
\cite{sachdev,kitaev,roberts},
I present one of the most basic ingredients of the theory, the moments $\{\mu_m\}$ \eqref{mu2mdef}
of a special operator $\eta$ called {\em sinusoidal coordinate} of many exactly solvable quantum 
systems in the Heisenberg picture.
They are ten discrete quantum mechanical systems of finite dimensions and two infinite ones plus
four ordinary one dimensional quantum mechanics.
In all these systems, the eigenvalues of the Hamiltonian are simple explicit functions of the 
system parameters and 
the sinusoidal coordinate is the variable in the eigenpolynomials 
$\{P_n(\eta)\}$ of the Hamiltonian $\mathcal{H}$.
These polynomials all belong to the Askey scheme of hypergeometric orthogonal polynomials 
\cite{askey, ismail, koeswart}.

This paper is organised as follows.
In section \ref{sec:set}, the basic concepts of Krylov complexity  are briefly recapitulated \cite{parker}
through the orthonormalisation of operators in a Krylov subspace \cite{krylov} by Lanczos algorithm \cite{lanczos}.
In section \ref{sec:solvquant} the outline of exactly solvable discrete quantum systems 
in the Schr\"odinger picture is briefly reviewed \cite{os12}.
Corresponding solutions of Heisenberg equation of motion \cite{os8,os7}, which are  the principal tool of the 
present paper, are discussed in some detail in section \ref{sec:Heisen}.
The main results of the paper, the exact and explicit expressions of the moments $\{\mu_m\}$ are derived
for ten exactly solvable discrete quantum mechanics in section \ref{sec:mu2m} and
four ordinary one-dimensional quantum mechanics in section \ref{sec:mumother}.
The inner product including the Boltzmann factor is introduced in section \ref{sec:mumother}
to deal with unbounded Hamiltonians.
The exact expressions of the moments for the bounded and unbounded Hamiltonians look
very similar as shown in two main  {\bf Theorems \ref{theo:main} and \ref{theo:mainb}}.
Among these 16 exactly solvable systems,   six systems related to the Krawtchouk and dual Hahn \S\ref{sec:KrdHn}, 
Meixner and Charlier \S\ref{sec:MeiChar}, Hermite  and Laguerre \S\ref{sec:HerLag}
polynomials 
show a very clear sign of `non-complexity' as vanishing higher Lanczos coefficients $b_m=0$, $m\ge3$.
These six systems  all share a linear spectrum $\mathcal{E}(n)\propto n$.
The final section is for a summary and some comments.

\section{Orthonormalisation of operators in the Krylov subspace}
\label{sec:set}

Let us start with a brief review of the general setting of  the Krylov complexity 
along the line of the seminal work \cite{parker} and others \cite{barbon,dymarsky,rabino,caputa} in order to introduce
proper notions and appropriate notation.
The orthonormalisation of operators in the Krylov subspace  is the basic ingredient for the evaluation of
Krylov complexity.
For simplicity of presentation,  the discrete quantum mechanical systems are discussed at the beginning.
That is, the basic vector space is assumed to be $\mathbb{C}^{N+1}$, $N\in\mathbb{N}$
in  this section and  up to \S\ref{sec:mu2m}. 
 Later in \S\ref{sec:mumother}, infinite dimensional Hilbert space is introduced for the treatment
 of the traditional  quantum mechanical systems.

Let us begin with the notation.
The ordinary Krylov subspace \cite{krylov} is spanned by a series of vectors generated by multiple 
applications of a certain operator $\mathcal{H}$ on a vector $v$,
\begin{equation*}
K_M(\mathcal{H};v)\eqdef\text{span}\{v,\mathcal{H}v,\mathcal{H}^2v,\ldots,\mathcal{H}^{M-1}v\}.
\end{equation*}
For the evaluation of Krylov complexity \cite{parker,barbon,dymarsky,rabino, caputa}, 
a different type of  subspaces is necessary.
It is spanned by a series of operators generated by multiple applications 
of adjoint actions of a Hamiltonian $\mathcal{H}$ 
on an operator, to be named as  $\eta$ throughout this paper,
\begin{equation}
\mathcal{K}_M(\mathcal{H};\eta)\eqdef
\text{span}\{\eta,\mathcal{L}\eta,\mathcal{L}^2\eta,\ldots,\mathcal{L}^{M-1}\eta\},
\label{Ksub}
\end{equation}
in which the Liouville operator $\mathcal{L}$ denotes the commutator by the Hamiltonian $\mathcal{H}$;
\begin{equation}
\mathcal{L}\eta\eqdef [\mathcal{H},\eta],\quad \mathcal{L}^2\eta=[\mathcal{H},\mathcal{L}\eta],\quad 
\cdots.
\label{liouville}
\end{equation}
The Hamiltonian  $\mathcal{H}$, $\eta$ and other 
operators  $\mathcal{V}$, $\mathcal{W}$ etc are all $(N+1)\times(N+1)$ matrices with the complex components,
\begin{equation}
\mathbb{C}\ni \mathcal{V}_{x\,y}, \mathcal{W}_{x\,y},\qquad x,y\in\mathcal{X}\eqdef\{0,1,\ldots,N\}.
\end{equation}
The Hamiltonian $\mathcal{H}$ is a {\em positive semi-definite hermitian} matrix,
\begin{equation}
\mathcal{H}^\dagger=\mathcal{H}, \quad \mathcal{H}^*_{x\,y}=\mathcal{H}_{y\,x},
\label{Hherm}
\end{equation}
in which $\dagger$ denotes  the hermitian conjugation and $*$ means the complex conjugation.
Throughout this paper the operator $\eta$ is assumed to be a real diagonal matrix,
\begin{equation}
\eta\eqdef\text{diag}\{\eta(0),\eta(1),\ldots,\eta(N)\},\quad \eta(x)\in\mathbb{R},\quad x\in\mathcal{X}.
\label{etadig}
\end{equation}

Two types of bra-ket notation are used. The bra $\lpar x|$ and ket $|x\rpar$ correspond to the $x$-th
unit vector $\bm{e}_x$  in $\mathbb{C}^{N+1}$, that is,
\begin{equation}
\lpar x|y\rpar=\delta_{x\,y},\qquad \lpar x|\mathcal{V}|y\rpar=\mathcal{V}_{x\,y},\quad x,y\in\mathcal{X}.
\label{pardef}
\end{equation}
The orthonormal basis corresponding to the eigenvectors of the Hamiltonian $\mathcal{H}$ is denoted by
the bra $\langle n|$ and the ket $|n\rangle$,
\begin{equation}
\langle m|n\rangle=\delta_{m\,n},\quad \langle m|\mathcal{H}=\langle m|\mathcal{E}(m),\quad
\mathcal{H}|n\rangle=\mathcal{E}(n)|n\rangle, \quad
\mathcal{E}(m),\mathcal{E}(n)\ge0,\quad m,n\in\mathcal{X}.
\end{equation}
The inner product of operators $\mathcal{V}$ and $\mathcal{W}$ is defined by the trace,
\begin{equation}
(\mathcal{V},\mathcal{W})\eqdef\text{Tr}\lbr\mathcal{V}^\dagger\mathcal{W}\rbr
=(\mathcal{W}^\dagger, \mathcal{V}^\dagger)=  
\sum_{x\in\mathcal{X}}\lpar x|\mathcal{V}^\dagger\mathcal{W}|x\rpar
=\sum_{n\in\mathcal{X}}\langle n|\mathcal{V}^\dagger\mathcal{W}|n\rangle=(\mathcal{W},\mathcal{V})^*,
\label{indef}
\end{equation}
which is real if $\mathcal{V}$ and $\mathcal{W}$ are both hermitian or anti-hermitian.
The norm of an operator $\mathcal{V}$ is denoted by,
\begin{equation}
\norm{\mathcal V}=\sqrt{(\mathcal{V},\mathcal{V})}
=\sqrt{\text{Tr}\lbr\mathcal{V}^\dagger\mathcal{V}\rbr}\ge0.
\label{normdef}
\end{equation}
It should be stressed that the Liouville operator $\mathcal{L}$ flips from the right to left side and vice versa,
under the present definition of the inner product,
\begin{equation}
(\mathcal{V},\mathcal{L}\mathcal{W})
=\text{Tr}\lbr\mathcal{V}^\dagger(\mathcal{H}\mathcal{W}-\mathcal{W}\mathcal{H})\rbr
=\text{Tr}\lbr(\mathcal{V}^\dagger\mathcal{H}-\mathcal{H}\mathcal{V}^\dagger)\mathcal{W}\rbr
=\text{Tr}\lbr[\mathcal{H},\mathcal{V}]^\dagger\mathcal{W}\rbr=(\mathcal{L}\mathcal{V},\mathcal{W}).
\label{Lflip}
\end{equation}
It is also obvious that $(\mathcal{V},\mathcal{L}\mathcal{V})$ vanishes if $\mathcal{V}$ 
is hermitian or anti-hermitian,
\begin{equation}
\mathcal{V}^\dagger=\pm\mathcal{V} \Longrightarrow (\mathcal{V},\mathcal{L}\mathcal{V})=\pm
\text{Tr}\lbr\mathcal{V}(\mathcal{H}\mathcal{V}-\mathcal{V}\mathcal{H})\rbr=0.
\label{vlv0}
\end{equation}

The orthonormalisation of the Krylov subspace $\mathcal{K}_m(\mathcal{H},\eta)$ \eqref{Ksub},
$\{\mathcal{O}_n\}$, $n=0,1,\ldots$
is much simpler than the ordinary Gram-Schmidt orthonormalisation due to the built-in structure of
$\mathcal{K}_M(\mathcal{H},\eta)$.
The orthonormalisation \`a la Lanczos \cite{lanczos}  starts with
\begin{align}
&\mathcal{O}_0\eqdef\eta/\norm{\eta},\qquad \norm{\eta}^2=\sum_{x\in\mathcal{X}}\eta(x)^2,
\label{ozerodef}\\
&\mathcal{W}_0\eqdef\mathcal{L}\mathcal{O}_0,
\quad b_1\eqdef\norm{\mathcal{W}_0}=\norm{\mathcal{L}\mathcal{O}_0},\quad 
\text{if} \ \ b_1\neq0 \ \Rightarrow \
\mathcal{O}_1\eqdef\frac1{b_1}\mathcal{W}_0,  
\label{step0}
\end{align}
and  it goes on until a zero norm operator is produced. 
By the above property \eqref{vlv0}, $\mathcal{O}_1$ 
is automatically orthogonal with the previous $\mathcal{O}_0$,
\begin{equation}
(\mathcal{O}_j,\mathcal{O}_l)=\delta_{j\,l},\quad j,l=0,1.
\label{10vert}
\end{equation}
The next step is the orthogonalisation of $\mathcal{L}\mathcal{O}_1$ with $\mathcal{O}_0$,
\begin{equation}
\mathcal{W}_1\eqdef\mathcal{L}\mathcal{O}_1-b_1\mathcal{O}_0,
\quad b_2\eqdef\norm{\mathcal{W}_1},\quad
\text{if} \ \ b_2\neq0 \ \Rightarrow \  
\mathcal{O}_2\eqdef\frac1{b_2}\mathcal{W}_1.
\label{step1}
\end{equation}
The orthogonality $(\mathcal{O}_1,\mathcal{W}_1)=0$ is obvious by \eqref{vlv0} and \eqref{10vert}.
By construction, the orthogonality $(\mathcal{O}_0,\mathcal{W}_1)=0$ holds. As
\begin{align*}
(\mathcal{O}_0,\mathcal{W}_1)&=(\mathcal{O}_0,\mathcal{L}\mathcal{O}_1)-b_1(\mathcal{O}_0,\mathcal{O}_0)
=(\mathcal{L}\mathcal{O}_0,\mathcal{O}_1)-b_1
=\frac1{b_1}(\mathcal{L}\mathcal{O}_0,\mathcal{L}\mathcal{O}_0)-b_1=0,
\end{align*}
the orthonormality up to 2 is established
\begin{equation}
(\mathcal{O}_j,\mathcal{O}_l)=\delta_{j\,l},\quad j,l=0,1,2.
\label{20vert}
\end{equation}
The flip property  $(\mathcal{V},\mathcal{L}\mathcal{W})=(\mathcal{L}\mathcal{V},\mathcal{W})$ \eqref{Lflip}
plays an important role in the above  and further calculations.
The process goes on as
\begin{equation}
\mathcal{W}_k\eqdef\mathcal{L}\mathcal{O}_k-b_k\mathcal{O}_{k-1},
\quad b_{k+1}\eqdef\norm{\mathcal{W}_k},\quad
\text{if} \ \ b_{k+1} \neq0 \ \Rightarrow \ 
\mathcal{O}_{k+1}\eqdef\frac1{b_{k+1}}\mathcal{W}_k.
\label{stepk}
\end{equation}
It is easy to prove that $\mathcal{O}_{k+1}$ is orthogonal to all the previous ones
$(\mathcal{O}_j,\mathcal{O}_{k+1})=0$, $j=0,1,\ldots,k$, by assuming the previous ones are
orthonormal
\begin{equation}
(\mathcal{O}_j,\mathcal{O}_l)=\delta_{j\,l},\quad j,l=0,1,\ldots,k.
\label{k0vert}
\end{equation}
It is obvious by construction $(\mathcal{O}_k,\mathcal{W}_{k})=0=(\mathcal{O}_k,\mathcal{O}_{k+1})$.
 Likewise,
\begin{align*}
(\mathcal{O}_{k-1},\mathcal{W}_{k})
&=\bigl(\mathcal{O}_{k-1},\mathcal{L}\mathcal{O}_{k}-b_k\mathcal{O}_{k}\bigr)
=\bigl(\mathcal{O}_{k-1},\mathcal{L}\mathcal{O}_{k}\bigr)-b_k
=\frac1{b_k}\bigl(\mathcal{O}_{k-1},\mathcal{L}\mathcal{W}_{k-1}\bigr)-b_k\\
&=\frac1{b_k}\bigl(\mathcal{L}\mathcal{O}_{k-1},\mathcal{W}_{k-1}\bigr)-b_k=
\frac1{b_k}\bigl(\mathcal{W}_{k-1}+b_{k-1}\mathcal{O}_{k-2},\mathcal{W}_{k-1}\bigr)-b_k=b_k-b_k=0,
\end{align*}
as $(\mathcal{O}_{k-2},\mathcal{W}_{k-1})=b_k(\mathcal{O}_{k-2},\mathcal{O}_{k})=0$ by assumption.
For $j\leq k-2$, the same logic goes
\begin{align*}
(\mathcal{O}_{j},\mathcal{W}_{k})
&=\bigl(\mathcal{O}_{j},\mathcal{L}\mathcal{O}_{k}-b_k\mathcal{O}_{k-1}\bigr)
=(\mathcal{O}_{j},\mathcal{L}\mathcal{O}_{k})=\frac1{b_k}(\mathcal{L}\mathcal{O}_{j},\mathcal{W}_{k-1})\\
&=\frac1{b_k}(\mathcal{W}_{j}+b_j\mathcal{O}_{j-1},\mathcal{W}_{k-1})=0,
\end{align*}
and the induction is complete. \hfill$\square$

\begin{rema}
\label{rema:stop}
The orthonormalisation is complete when all the Lanczos coefficients $\{b_n\}$ are determined.
The orthonormalisation stops at $\mathcal{O}_k$ when $b_{k+1}$ vanishes, $b_{k+1}=0$.  
Two explicit examples of the stopped orthonormalisation will be shown in \S\ref{sec:KrdHn}.
Since the Hilbert space is $\mathbb{C}^{N+1}$, the totality of the basis $\{\mathcal{O}_n\}$ 
 is less than $(N+1)^2$.
\end{rema}
\begin{rema}
\label{rema:herm}
The orthonormal basis $\mathcal{O}_n$ has the following structure
\begin{equation}
\mathcal{O}_n=\sum_{j=0}^{[\tfrac n2]}c_j^{(n)}\mathcal{L}^{n-2j}\mathcal{O}_0,\quad 
\bigl(i^n\mathcal{O}_n\bigr)^\dagger=i^n\mathcal{O}_n,
\label{onstruc}
\end{equation}
in which $[m]$ is the Gauss's symbol meaning the greatest integer not exceeding $m$.
The squares of the  Lanczos coefficients up to $n$, $\{b_1^2,\ldots,b_n^2\}$, which are the length squared of the
basis before normalisation, are expressed as rational functions of the moments $\mu_{2m}$,
\begin{equation}
\mu_{2m}\eqdef (\mathcal{O}_0,\mathcal{L}^{2m}\mathcal{O}_0),\quad 0\leq m\le n,\quad \mu_0=1.
\label{mu2mdef}
\end{equation}
For example {\rm (cf  \cite{vis} Table 3-2)},
\begin{equation}
b_1^2=\mu_2,\quad b_2^2=\frac{\mu_4}{\mu_2}-\mu_2,\quad
b_3^2=\frac{\mu_2(\mu_6-2\mu_2\mu_4+\mu_2^3)}{\mu_2(\mu_4-\mu_2^2)}-\frac{\mu_4}{\mu_2}+\mu_2.
\label{b123}
\end{equation}
Such formulas can be checked by considering the formal scaling properties,
\begin{equation}
\mathcal{H}\to\lambda\mathcal{H}\ \Longrightarrow \mu_{2m}\to\lambda^{2m}\mu_{2m},\quad
b_n\to\lambda b_n.
\label{scaling}
\end{equation}
\end{rema}
\begin{rema}
\label{rema:hankel}
In some work {\rm \cite{parker}(A.4)},  
a formula involving the   determinant of Hankel matrix of moments
\begin{equation*}
b_1^2\cdots b_n^2={\rm det}(\mu_{i+j})_{0\leq i,j\leq n}
\end{equation*}
was erroneously reported.  The l.h.s. scales as $\lambda^{2n}$ and the  diagonal part of the r.h.s.
matrix,  
$1\cdot\mu_2\cdots\mu_{2n}$,  scales as $\lambda^{n(n+1)}$.
\end{rema}
\begin{rema} 
\label{rema:vis}
In  a monograph {\rm \cite{vis}} Viswanath and  M\"uller reported 
a recursive formula to determine $\{b_1^2,\ldots,b_K^2\}$
based on the knowledge of $\mu_2,\ldots,\mu_{2K}$  {\rm \cite{vis}(3.33)} and its reverse,
from  $\mu_2,\ldots,\mu_{2K}$ to $\{b_1^2,\ldots,b_K^2\}$,  {\rm \cite{vis}(3.34)}.  
These formulas were recapitulated in {\rm \cite{parker}(A.5),(A.7)}.
\end{rema}
\begin{defi}
\label{def:kcomp}
Krylov complexity $K(\mathcal{H},\eta;t)$ is defined 
{\rm \cite{parker,barbon,rabino,caputa}} based on the Heisenberg operator
solution of $\mathcal{O}(t)$ of $\eta$ and its projection component $\varphi_n(t)$ on $\mathcal{O}_n$,
\begin{equation}
K(\mathcal{H},\eta;t)\eqdef \sum_nn\varphi_n(t)^2,
\quad \varphi_n(t)\eqdef\bigl(i^n\mathcal{O}_n,\mathcal{O}(t)\bigr)\in\mathbb{R}, \quad n=1,2,\ldots,
\label{Ketdef}
\end{equation}
in which $\mathcal{O}(t)$ is the Heisenberg operator solution of $\eta$,
\begin{equation}
\mathcal{O}(t)\eqdef e^{i\mathcal{H}t}\mathcal{O}_0e^{-i\mathcal{H}t}
=e^{i\mathcal{H}t}\eta e^{-i\mathcal{H}t}/\norm{\eta},\quad \bigl(\mathcal{O}(t),\mathcal{O}(t)\bigr)=1.
\label{Heisensol}
\end{equation}
Reflecting the unit norm of $\mathcal{O}(t)$, $\sum_n\varphi(t)^2=1$ holds.
\end{defi}

\bigskip
In the rest of this paper I present 
the {\em explicit forms of the Heisenberg operator solution \eqref{Heisensol} and the moments $\{\mu_{2m}\}$}
of more than a dozen quantum mechanical systems. 
Based on the exact knowledge of the moments, the Lanczos coefficients $\{b_n\}$ and the functions 
$\{\varphi_n(t)\}$ \eqref{Ketdef} can be evaluated as {\em precisely as wanted}.

\section{Exactly solvable discrete quantum mechanics}
\label{sec:solvquant}
Here I present ten exactly solvable discrete quantum mechanical systems, see \cite{os24} for a review. 
The eigenvectors of the Hamiltonians are the hypergeometric orthogonal polynomials of the Askey scheme
\cite{askey,ismail,koeswart,os12}.
They are the Krawtchouk (K), Hahn (H), dual Han (dH), Racah (R), 
quantum $q$-Krawtchouk (q$q$K), $q$-Krawtchouk ($q$K), affine $q$-Krawtchouk (a$q$K),
$q$-Hahn ($q$H), dual $q$-Hahn (d$q$H) and $q$-Racah ($q$R) polynomials.
The Hamiltonian $\mathcal{H}$ of these exactly solvable quantum mechanics   is 
a {\em tridiagonal}  $(N+1)\times(N+1)$ real symmetric matrix,
\begin{align}
\hspace{-2mm}   \mathcal{H}_{x\,y}=
  \bigl(B(x)+D(x)\bigr)\delta_{x,y}
 \! -\!\sqrt{B(x)D(x+1)}\,\delta_{x+1,y}\!-\!\sqrt{B(x-1)D(x)}\,\delta_{x-1,y},\ x,y\in\mathcal{X},
  \label{Hdef}  
\end{align}
{\scriptsize
\begin{equation*}
\mathcal{H}=\left(
\begin{array}{cccccc}
\!\!B(0)  & -\sqrt{B(0)D(1)}  &   0& \cdots&\cdots&0\\
\!\!-\sqrt{B(0)D(1)}  & B(1)+D(1)  & -\sqrt{B(1)D(2)} &0&\cdots&\vdots  \\
\!\!0  &  -\sqrt{B(1)D(2)}  &   B(2)+D(2)&-\sqrt{B(2)D(3)}&\cdots&\vdots\\
\!\!\vdots&\cdots&\cdots&\cdots&\cdots&\vdots\\
\!\!\vdots&\cdots&\cdots&\cdots&\cdots&0\\
\!\!0&\cdots&\cdots&-\sqrt{\!B(N\!\!-\!\!2)D(N\!\!-\!\!1)}&B(N\!\!-\!\!1)\!
+\!D(N\!\!-\!\!1)&-\sqrt{\!B(N\!\!-\!\!1)D(N)}\\
\!\!0&\cdots&\cdots&0&-\sqrt{B(N\!-\!1)D(N)}&D(N)
\end{array}
\right).
\end{equation*}
}
in which the functions $B(x)$ and $D(x)$ are {\em positive} except for  the boundary conditions,
\begin{equation}
B(x), D(x)>0,\quad D(0)=0,\quad B(N)=0.
\label{bdcon}
\end{equation}
The orthonormal eigenvectors of the Hamiltonian $\mathcal{H}$ are
\begin{align}
&\mathcal{H}|n\rangle=\mathcal{E}(n)|n\rangle,\quad \mathcal{E}(n)\ge0,\quad n\in\mathcal{X},
\label{ndef1}\\
&
\lpar x|n\rangle=\phi_0(x)P_n(\eta)d_n,\quad
 \sum_{x\in\mathcal{X}}\phi_0^2(x)P_m(\eta)P_n(\eta)=\frac{\delta_{m\,n}}{d_n^2} \Leftrightarrow 
\  \langle m|n\rangle=\delta_{m\,n},
\label{ndef2}
\end{align}
in which $P_n(\eta)$ is a degree $n$ polynomial in $\eta$.
The {\em sinusoidal coordinate} $\eta$ is a linear or quadratic function of $x$ or $q^{\pm x}$ ($0<q<1$) which
vanishes at $x=0$, $\eta(0)=0$ \cite{os12},
\begin{equation}
   \begin{split}
  &\text{(\romannumeral1)}\ \eta(x)=x,\  K, H,\quad  \text{(\romannumeral2)}\ \eta(x)=x(x+d),\  dH, R, \\
&\text{(\romannumeral3)}\ \eta(x)=1-q^x, \qquad \text{(\romannumeral4)}
\ \eta(x)=q^{-x}-1,\ qH, qqK, qK, aqK, \\
&\text{(\romannumeral5)}\ \eta(x)=(q^{-x}-1)(1-d q^x),\quad dqH, qR.
  \end{split}
  \label{5eta}
\end{equation}
Likewise, the eigenvalue $\mathcal{E}$ is a linear or quadratic function of $n$ or $q^{\pm n}$ which
vanishes at $n=0$, $\mathcal{E}(0)=0$ \cite{os12},
\begin{equation}
   \begin{split}
  &\text{(\romannumeral1)}\ \mathcal{E}(n)=n,\  K, dH,
  \quad  \text{(\romannumeral2)}\ \mathcal{E}(n)=n(n+d),\  H, R, \\
&\text{(\romannumeral3)}\ \mathcal{E}(n)=1-q^n,\ qqK, \qquad \text{(\romannumeral4)}
\ \mathcal{E}(n)=q^{-n}-1,\ dqH, aqK, \\
&\text{(\romannumeral5)}\ \mathcal{E}(n)=(q^{-n}-1)(1-d q^n),\quad qK, qH, qR.
  \end{split}
  \label{En}
\end{equation}  
In the formulas \eqref{5eta}, \eqref{En} the parameter $d$ is specific in each system.
As functions of $x$, $\{P_n\bigl(\eta(x)\bigr)\}$ are terminating ($q$)-hypergeometric functions 
\cite{askey,ismail,koeswart} and they are normalised by a uniform condition,
\begin{equation}
P_n(0)=1,\quad n\in\mathcal{X}.
\label{unicon}
\end{equation}
As  orthogonal polynomials in $\eta$ with the above normalisation condition, 
$\{P_n(\eta)\}$ satisfy {\em three term recurrence relation} \cite{os12},
\begin{align}
  \eta P_n(\eta)&=A_n\bigl(P_{n+1}(\eta)-P_{n}(\eta)\bigr)
  +C_n\bigl(P_{n-1}(\eta)-P_{n}(\eta)\bigr),
  \label{3term}
  \end{align}
in which the coefficients $A_n$, $C_n$ are {\em negative} except for  the boundary conditions
\begin{equation}
A_n,\ C_n<0,\quad C_0=0,\quad A_N=0,
\label{bdcon2}
\end{equation}
so that $P_{-1}$ and $P_{N+1}$ do not enter into the theory.

For definiteness, I show the data of the simplest example, the Krawtchouk (K) system:
\begin{align*}
  &B(x)=p(N-x),\quad
  D(x)=(1-p)x,\qquad 0<p<1,\\
  &P_n(\eta)=P_n(x)
  ={}_2F_1\Bigl(
  \genfrac{}{}{0pt}{}{-n,\,-x}{-N}\Bigm|p^{-1}\Bigr),\quad  \mathcal{E}(n)=n,\quad   \eta(x)=x,\\
  &\phi_0^2(x)=
  \frac{N!}{x!\,(N-x)!}\Bigl(\frac{p}{1-p}\Bigr)^x,\quad
d_n^2
  =\frac{N!}{n!\,(N-n)!}\Bigl(\frac{p}{1-p}\Bigr)^n\times(1-p)^N,\\
  &A_n=-p(N-n),\quad
  C_n=-(1-p)n.
\end{align*}
As will be shown shortly, most of the data, except for $\eta$, $\mathcal{E}$, $A_n$ and $C_n$, 
are not needed for the  evaluation of  the moments $\{\mu_{2m}\}$ and the functions $\{\varphi_n(t)\}$.

%
%
\section{Solutions of  Heisenberg equation of motion }
\label{sec:Heisen}
As shown in \eqref{Heisensol}, the explicit form of the Heisenberg operator solution is essential
for the determination of the functions $\{\varphi_n(t)\}$.
Although the Heisenberg solution of the harmonic oscillator potential ($x^2$) 
was known in the early days of quantum mechanics,
it was late 1970's that those for four other potentials in one-dimensional quantum mechanics
were reported by Nieto and Simmons \cite{nieto1}--\cite{nieto4}. 
They were for the potentials $x^2+1/x^2$, $1/\sin^2x$, $-1/\cosh^2x$
and the Morse potential.
The term `sinusoidal coordinate' was also coined by them, 
meaning that $\eta$ undergoes sinusoidal motion with frequencies depending on the energy.
About a quarter century after Nieto and Simmons, the list of exact Heisenberg operator solutions 
was enlarged by Odake and myself to include many discrete quantum mechanics \cite{os12,os8,os7,os13}
and some multi-particle dynamics \cite{os9}.

The essence is the discovery \cite{os7} (this paper will be cited as I) that the Hamiltonian 
$\mathcal{H}$ and the sinusoidal coordinate $\eta$
of  exactly sovable systems in the Schr\"odinger picture all satisfy a simple commutation relation
\begin{equation}
  [\mathcal{H},[\mathcal{H},\eta]\,]=\eta\,R_0(\mathcal{H})
  +[\mathcal{H},\eta]\,R_1(\mathcal{H})+R_{-1}(\mathcal{H}).
  \label{twocom}
\end{equation}
in which $R_0(\mathcal{H})$, $R_1(\mathcal{H})$ and $R_{-1}(\mathcal{H})$ are polynomials
in $\mathcal{H}$ of maximal degree 2, 1 and 2, respectively, 
reflecting the power counting of $\mathcal{H}$ on both sides.
Those $R_i(\mathcal{H})$'s may contain some system parameters but 
not dynamical operators other than $\mathcal{H}$.
In terms of the Liouville operator $\mathcal{L}$, the above commutation relation reads 
\begin{equation}
  \mathcal{L}^2\eta=\eta\,R_0(\mathcal{H})
  + \mathcal{L}\eta\,R_1(\mathcal{H})+R_{-1}(\mathcal{H}),
\label{Ltwocom}
\end{equation}
which is obviously generalised to
\begin{equation*}
  \mathcal{L}^m\eta=\eta\,\mathcal{A}_m(\mathcal{H})
  + \mathcal{L}\eta\,\mathcal{B}_m(\mathcal{H})
  +\mathcal{C}_m(\mathcal{H}),\quad m\in\mathbb{Z}_{\ge0},
\end{equation*}
with the obvious initial conditions
\begin{equation}
  \begin{split}
&   \mathcal{A}_0(\mathcal{H})=1,\quad \mathcal{A}_1(\mathcal{H})=0, 
\quad \mathcal{A}_2(\mathcal{H})=R_0(\mathcal{H}),\\
&  \mathcal{B}_0(\mathcal{H})=0,\quad \ \mathcal{B}_1(\mathcal{H})=1, 
\quad \mathcal{B}_2(\mathcal{H})=R_1(\mathcal{H}),\\
&  \mathcal{C}_0(\mathcal{H})=0,\quad \ \, \mathcal{C}_1(\mathcal{H})=0, 
\quad \  \mathcal{C}_2(\mathcal{H})=R_{-1}(\mathcal{H}).
   \end{split}
   \label{ABCini}
\end{equation}
By solving the recurrence relations,
\begin{align*}
&\mathcal{A}_{m+1}(\mathcal{H})=R_0(\mathcal{H})\mathcal{B}_m(\mathcal{H}),\qquad
\mathcal{B}_{m+1}(\mathcal{H})
=\mathcal{A}_{m}(\mathcal{H})+R_1(\mathcal{H})\mathcal{B}_m(\mathcal{H}),\\
&\mathcal{C}_{m+1}(\mathcal{H})=R_{-1}(\mathcal{H})\mathcal{B}_m(\mathcal{H})
=\frac{R_{-1}(\mathcal{H})}{R_0(\mathcal{H})}\mathcal{A}_{m+1}(\mathcal{H}),
\end{align*}
one arrives at,
\begin{align}
 \mathcal{L}^m\eta=&\,\eta\,\mathcal{A}_m(\mathcal{H})
  + \mathcal{L}\eta\,\mathcal{B}_m(\mathcal{H})
  +\mathcal{C}_m(\mathcal{H}),\qquad m\in\mathbb{Z}_{\ge0},
\label{Lmcom}\\
&\mathcal{A}_m(\mathcal{H})
=R_{0}(\mathcal{H})\frac{\alpha_+(\mathcal{H})^{m-1}-\alpha_-(\mathcal{H})^{m-1}}
{\alpha_+(\mathcal{H})-\alpha_-(\mathcal{H})},
\label{amf}\\
&\mathcal{B}_m(\mathcal{H})
=\frac{\alpha_+(\mathcal{H})^m-\alpha_-(\mathcal{H})^m}{\alpha_+(\mathcal{H})-\alpha_-(\mathcal{H})},
\label{bmf}\\
&\mathcal{C}_m(\mathcal{H})
=R_{-1}(\mathcal{H})\frac{\alpha_+(\mathcal{H})^{m-1}-\alpha_-(\mathcal{H})^{m-1}}
{\alpha_+(\mathcal{H})-\alpha_-(\mathcal{H})},\quad m\ge1,
\label{cmf}
\end{align}
in which two operators $\alpha_\pm(\mathcal{H})$ are the roots of the quadratic equation,
\begin{align}
&\alpha(\mathcal{H})^2-R_1(\mathcal{H})\alpha(\mathcal{H})-R_0(\mathcal{H})=0,
\label{aleq}\\
& \alpha_\pm(\mathcal{H})
=\frac12\left(R_1(\mathcal{H})\pm \sqrt{R_1(\mathcal{H})^2+4R_0(\mathcal{H})}\right),
\label{alpm}\\
&  \alpha_+(\mathcal{H})+\alpha_-(\mathcal{H})=R_1(\mathcal{H}),
  \quad
  \alpha_+(\mathcal{H})\alpha_-(\mathcal{H})=-R_0(\mathcal{H}).
  \label{al+x}
\end{align}
They also satisfy the relations as shown in \cite{os7}(I.2.22) and (I.2.23)
\begin{align}
 & \mathcal{E}(n+1)-\mathcal{E}(n)=\alpha_+\bigl(\mathcal{E}(n)\bigr),\quad
  \mathcal{E}(n-1)-\mathcal{E}(n)=\alpha_-\bigl(\mathcal{E}(n)\bigr).
\label{alpm1}\\
 &  \mathcal{E}(n)-\mathcal{E}(n-1)=\alpha_+\bigl(\mathcal{E}(n-1)\bigr),\quad
  \mathcal{E}(n)-\mathcal{E}(n+1)=\alpha_-\bigl(\mathcal{E}(n+1)\bigr),
\label{alpm2}\\
&\alpha_+\bigl(\mathcal{E}(n-1)\bigr)
=-\alpha_-\bigl(\mathcal{E}(n)\bigr),\quad \alpha_-\bigl(\mathcal{E}(n+1)\bigr)
=-\alpha_+\bigl(\mathcal{E}(n)\bigr).
\label{alpm3}
\end{align}
Summing up $\{\mathcal{L}^m\eta\}$ leads to the exact Heisenberg operator solution of $\eta$ \cite{os8,os7},
\begin{align}
   e^{it\mathcal{H}}\eta\,e^{-it\mathcal{H}}
  &=\sum_{m=0}^\infty\frac{(it)^m}{m!}\mathcal{L}^m\eta,\n
  &=\mathcal{L}\eta\cdot
  \frac{e^{i\alpha_+(\mathcal{H})t}-e^{i\alpha_-(\mathcal{H})t}}
  {\alpha_+(\mathcal{H})-\alpha_-(\mathcal{H})}
  -R_{-1}(\mathcal{H})/R_{0}(\mathcal{H})\n
  &\quad
  +\bigl(\eta+R_{-1}(\mathcal{H})/R_0(\mathcal{H})\bigr)
  \frac{-\alpha_-(\mathcal{H})e^{i\alpha_+(\mathcal{H})t}
  +\alpha_+(\mathcal{H})e^{i\alpha_-(\mathcal{H})t}}
  {\alpha_+(\mathcal{H})-\alpha_-(\mathcal{H})}.
  \label{quantsol}
\end{align}
Strictly speaking, $\eta$ on both sides represents the initial value at $t=0$ of the time-dependent Heisenberg
operator.
It should be stressed that the derivation from \eqref{twocom} to \eqref{Lmcom}--\eqref{cmf} 
and \eqref{quantsol} is purely algebraic. It is valid in any quantum system, not necessarily finite dimensional ones.
As shown above, $\eta$'s time development is a superposition of two sinusoidal waves of frequencies 
$\alpha_\pm\bigl(\mathcal{H}\bigr)$ depending on the total energy.

A simple and useful information is extracted from the above expression \eqref{quantsol}, 
the energy level dependence of the expectation value of the Heisenberg operator 
$e^{it\mathcal{H}}\eta\,e^{-it\mathcal{H}}$, which is obviously time-independent
$\langle n|e^{it\mathcal{H}}\eta\,e^{-it\mathcal{H}}|n\rangle=\langle n|\eta|n\rangle$.
Sandwiching \eqref{quantsol} by $\langle n|$ and $|n\rangle$ and noting $\langle n|\mathcal{L}\eta|n\rangle=0$,
one obtains
\begin{align}
&\left(\langle n|\eta|n\rangle+\frac{R_{-1}\bigl(\mathcal{E}(n)\bigr)}{R_{0}\bigl(\mathcal{E}(n)\bigr)}\right)
\times\left(1-\frac{-\alpha_-\bigl(\mathcal{E}(n)\bigr)e^{i\alpha_+\left(\mathcal{E}(n)\right)t}
  +\alpha_+\bigl(\mathcal{E}(n)\bigr)e^{i\alpha_-\left(\mathcal{E}(n)\right)t}}
  {\alpha_+\bigl(\mathcal{E}(n)\bigr)-\alpha_-\bigl(\mathcal{E}(n)\bigr)}\right)=0,\n
& \hspace{40mm} \Longrightarrow \langle n|\eta|n\rangle
=-\frac{R_{-1}\bigl(\mathcal{E}(n)\bigr)}{R_{0}\bigl(\mathcal{E}(n)\bigr)},\qquad n\in\mathcal{X},
\label{netn}
\end{align}
as the second factor is time-dependent and non-vanishing.
\section{Evaluation of $\mu_{2m}$ of exactly solvable discrete quantum mechanics}
\label{sec:mu2m}
With the formulas \eqref{Lmcom}-\eqref{cmf} and \eqref{netn}, the derivation of the exact
expressions of the moment $\mu_m$ \eqref{mu2mdef} is straightforward.
For the evaluation of moment ($m\ge1$),
\begin{align}
\mu_m&=\sum_{n\in\mathcal{X}}\langle n|\mathcal{O}_0\mathcal{L}^m\mathcal{O}_0|n\rangle
=\sum_{n\in\mathcal{X}}\langle n|\eta\mathcal{L}^m\eta|n\rangle/\norm{\eta}^2\n
&=\sum_{n\in\mathcal{X}}\left\{\langle n|\eta^2|n\rangle\langle n|\mathcal{A}_m(\mathcal{H})|n\rangle
+\langle n|\eta\mathcal{L}\eta|n\rangle\langle n|\mathcal{B}_m(\mathcal{H})|n\rangle
+\langle n|\eta|n\rangle\langle n|\mathcal{C}_m(\mathcal{H})|n\rangle\right\}/\norm{\eta}^2,
\label{mum1}
\end{align}
 all the coefficient terms $\langle n|\mathcal{A}_m(\mathcal{H})|n\rangle$, 
$\langle n|\mathcal{B}_m(\mathcal{H})|n\rangle$, $\langle n|\mathcal{C}_m(\mathcal{H})|n\rangle$ are known
\begin{equation*}
\mathcal{H}|n\rangle=\mathcal{E}(n)|n\rangle \Rightarrow \
\langle n|\alpha_\pm(\mathcal{H})|n\rangle=\alpha_\pm\bigl(\mathcal{E}(n)\bigr),
\quad \langle n|R_i(\mathcal{H})|n\rangle=R_i\bigl(\mathcal{E}(n)\bigr), \quad i=-1,0,1.
\end{equation*}
One only needs the values of 
\begin{equation*}
\langle n|\eta^2|n\rangle, \quad \langle n|\eta\mathcal{L}\eta|n\rangle, \quad \langle n|\eta|n\rangle,
\end{equation*}
and they are expressed by the coefficients $A_n$ and $C_n$ of the three term recurrence relation \eqref{3term}.
By using the explicit expression $\lpar x|n\rangle=\phi_0(x)P_n(\eta)d_n$ and the three term recurrence 
relation, one obtains
\begin{align}
\eta|n\rangle=-(A_n+C_n)|n\rangle +A_n\frac{d_n}{d_{n+1}}|n+1\rangle
+C_n\frac{d_n}{d_{n-1}}|n-1\rangle,\n
 \langle n|\eta|n\rangle=-(A_n+C_n)
 =-\frac{R_{-1}\bigl(\mathcal{E}(n)\bigr)}{R_{0}\bigl(\mathcal{E}(n)\bigr)},
\label{etan}\\
 \langle n+1|\eta|n\rangle= \langle n|\eta|n+1\rangle=A_n\frac{d_n}{d_{n+1}},
\quad
\langle n-1|\eta|n\rangle=\langle n|\eta|n-1\rangle=C_n\frac{d_n}{d_{n-1}}.
\label{npmetan}\\
\eta|n+1\rangle=-(A_{n+1}+C_{n+1})|n+1\rangle +A_{n+1}\frac{d_{n+1}}{d_{n+2}}|n+2\rangle
+C_{n+1}\frac{d_{n+1}}{d_{n}}|n\rangle,\n
 \langle n|\eta|n+1\rangle=C_{n+1}\frac{d_{n+1}}{d_{n}},
\label{npeta}\\
\eta|n-1\rangle=-(A_{n-1}+C_{n-1})|n-1\rangle +A_{n-1}\frac{d_{n-1}}{d_{n}}|n\rangle
+C_{n-1}\frac{d_{n-1}}{d_{n-2}}|n-2\rangle,\n
\langle n|\eta|n-1\rangle=A_{n-1}\frac{d_{n-1}}{d_{n}},
\label{npeta1}\\
\Longrightarrow \quad  \langle n|\eta|n+1\rangle^2=A_nC_{n+1},\quad n\le N-1,
\quad \langle n|\eta|n-1\rangle^2=A_{n-1}C_{n},
\quad n\ge1.
 \label{npmform}
\end{align}
By combining them, one arrives at
\begin{align}
\langle n|\eta^2|n\rangle&=\sum_{\ell=n,n\pm1}\langle n|\eta|\ell\rangle\langle\ell|\eta|n\rangle
= \langle n|\eta|n\rangle^2+ \langle n|\eta|n+1\rangle^2+\langle n|\eta|n-1\rangle^2\n
&=(A_n+C_n)^2+A_nC_{n+1}+A_{n-1}C_n,
\label{neta2n}\\
\langle n|\eta\mathcal{L}\eta|n\rangle&=\langle n|\eta\mathcal{H}\eta|n\rangle
-\langle n|\eta^2\mathcal{H}|n\rangle\n
&=\sum_{\ell=n,n\pm1}\langle n|\eta|\ell\rangle\mathcal{E}(\ell)\langle \ell|\eta|n\rangle
-\mathcal{E}(n)\langle n|\eta^2|n\rangle\n
&=(A_n+C_n)^2\mathcal{E}(n)+A_nC_{n+1}\mathcal{E}(n+1)+A_{n-1}C_n\mathcal{E}(n-1)\n
& \quad -\mathcal{E}(n)\Bigl((A_n+C_n)^2+A_nC_{n+1}+A_{n-1}C_n\Bigr)\n
&=A_nC_{n+1}\,\alpha_+\bigl(\mathcal{E}(n)\bigr)+A_{n-1}C_{n}\,\alpha_-\bigl(\mathcal{E}(n)\bigr),
\end{align}
by using \eqref{alpm1}. It is easy to see that $(A_n+C_n)^2\mathcal{A}_m\bigl(\mathcal{E}(n)\bigr)$
cancels $\langle n|\eta|n\rangle\mathcal{C}_m\bigl(\mathcal{E}(n)\bigr)$
 in \eqref{mum1} due to \eqref{netn}.
These lead to
\begin{align*}
\norm{\eta}^2\mu_m&=\ \sum_{n=0}^{N-1}A_nC_{n+1}\Bigl(A_m\bigl(\mathcal{E}(n)\bigr)
+\alpha_+\bigl(\mathcal{E}(n)\bigr)B_m\bigl(\mathcal{E}(n)\bigr)\Bigr)\\
&\ \ +\sum_{n=1}^NA_{n-1}C_{n}\Bigl(A_m\bigl(\mathcal{E}(n)\bigr)
+\alpha_-\bigl(\mathcal{E}(n)\bigr)B_m\bigl(\mathcal{E}(n)\bigr)\Bigr).
\end{align*}
With $R_0\bigl((\mathcal{E}(n)\bigr)=-\alpha_+\bigl(\mathcal{E}(n)\bigr)\alpha_-\bigl(\mathcal{E}(n)\bigr)$
\eqref{al+x},
\begin{align}
&\mathcal{A}_{m}\bigl(\mathcal{E}(n)\bigr)
+\alpha_+\bigl(\mathcal{E}(n)\bigr)\mathcal{B}_{m}\bigl(\mathcal{E}(n)\bigr)\n
&=-\alpha_+\bigl(\mathcal{E}(n)\bigr)\alpha_-\bigl(\mathcal{E}(n)\bigr)
\frac{\alpha_+\bigl(\mathcal{E}(n)\bigr)^{m-1}-\alpha_-\bigl(\mathcal{E}(n)\bigr)^{m-1}}
{\alpha_+\bigl(\mathcal{E}(n)\bigr)-\alpha_-\bigl(\mathcal{E}(n)\bigr)}\n
&\qquad +\alpha_+\bigl(\mathcal{E}(n)\bigr)
\frac{\alpha_+\bigl(\mathcal{E}(n)\bigr)^{m}-\alpha_-\bigl(\mathcal{E}(n)\bigr)^{m}}
{\alpha_+\bigl(\mathcal{E}(n)\bigr)-\alpha_-\bigl(\mathcal{E}(n)\bigr)}
=\alpha_+\bigl(\mathcal{E}(n)\bigr)^{m},
\label{plussum}
\end{align}
and
\begin{align}
&\mathcal{A}_{m}\bigl(\mathcal{E}(n)\bigr)
+\alpha_-\bigl(\mathcal{E}(n)\bigr)\mathcal{B}_{m}\bigl(\mathcal{E}(n)\bigr)\n
&=-\alpha_+\bigl(\mathcal{E}(n)\bigr)\alpha_-\bigl(\mathcal{E}(n)\bigr)
\frac{\alpha_+\bigl(\mathcal{E}(n)\bigr)^{m-1}-\alpha_-\bigl(\mathcal{E}(n)\bigr)^{m-1}}
{\alpha_+\bigl(\mathcal{E}(n)\bigr)-\alpha_-\bigl(\mathcal{E}(n)\bigr)}\n
&\qquad +\alpha_-\bigl(\mathcal{E}(n)\bigr)
\frac{\alpha_+\bigl(\mathcal{E}(n)\bigr)^{m}-\alpha_-\bigl(\mathcal{E}(n)\bigr)^{m}}
{\alpha_+\bigl(\mathcal{E}(n)\bigr)-\alpha_-\bigl(\mathcal{E}(n)\bigr)}
=\alpha_-\bigl(\mathcal{E}(n)\bigr)^{m},
\label{minsum}
\end{align}
one arrives at
\begin{align*}
\norm{\eta}^2\mu_m&
=\sum_{n=0}^{N-1}A_nC_{n+1}\Bigl(\alpha_+\bigl(\mathcal{E}(n)\bigr)\Bigr)^{m}
+\sum_{n=1}^{N}A_{n-1}C_{n}\Bigl(\alpha_-\bigl(\mathcal{E}(n)\bigr)\Bigr)^{m},
\end{align*}
in which the boundary conditions \eqref{bdcon2} $A_N=0$ ans $C_0=0$ are used.
Changing $n-1\to n$ in the second sum leads to 
\begin{align*}
\norm{\eta}^2\mu_m&=\sum_{n=0}^{N-1}A_nC_{n+1}\Bigl(\alpha_+\bigl(\mathcal{E}(n)\bigr)^{m}
+\alpha_-\bigl(\mathcal{E}(n+1)\bigr)^{m}\Bigr)\n
&=\sum_{n=0}^{N-1}A_nC_{n+1}\Bigl(\alpha_+\bigl(\mathcal{E}(n)\bigr)^{m}
+\bigl(-\alpha_+\bigl(\mathcal{E}(n)\bigr)\bigr)^{m}\Bigr)\\
&=2\sum_{n=0}^{N-1}A_nC_{n+1}\Bigl(\alpha_+\bigl(\mathcal{E}(n)\bigr)\Bigr)^{m}
\times
\left\{
\begin{array}{cc}
1  &   m:\ \text{even}   \\
0  &   m:\ \text{odd}   
\end{array}
\right.
\label{finmum},
\end{align*}
in which \eqref{alpm3} is used. 
\begin{rema}
\label{normindep}
It should be stressed that the product of the coefficients of the three term recurrence relations \eqref{3term}
$A_nC_{n+1}$ {\rm (}$A_{n-1}C_{n}${\rm )} is independent of the
normalisation change of the polynomial 
$P_n(\eta)\to P_n(\eta)'=\gamma_nP_n(\eta)$, $n\in\mathbb{Z}_{\ge0}$.
$\gamma_n\neq0$.
\end{rema}
These results are summarised as  the following
\begin{theo}
\label{theo:main}
The moments of the exactly solvable discrete quantum systems have a very simple exact expression
\begin{equation}
\mu_{2m}=2\sum_{n=0}^{N-1}A_nC_{n+1}
\left(\alpha_+\bigl(\mathcal{E}(n)\bigr)\right)^{2m}/\norm{\eta}^2,
\quad \mu_{2m-1}=0,\quad m\in\mathbb{N}.
\label{mainth}
\end{equation}
It applies to ten systems related to the Krawtchouk (K), Hahn (H), dual Han (dH), Racah (R), 
quantum $q$-Krawtchouk (q$q$K), $q$-Krawtchouk ($q$K), affine $q$-Krawtchouk (a$q$K),
$q$-Hahn ($q$H), dual $q$-Hahn (d$q$H) and $q$-Racah ($q$R) polynomials.
\end{theo}

In the rest of this section, the necessary data for the evaluation of the moments $\mu_{2m}$ of 
the ten discrete quantum systems are provided.
They are $B(x)$ and $D(x)$ for the definition of the Hamiltonian 
$\mathcal{H}$ and for specifying the parameter ranges,
the sinusoidal coordinates $\eta(x)$, the energy eigenvalues $\mathcal{E}(n)$, $R_0(\mathcal{H})$ and
$R_1(\mathcal{H})$ for the  derivation of $\alpha_\pm\bigl(\mathcal{E}(n)\bigr)$ and the coefficients of the
three term recurrence relation $A_n$ and $C_n$.
It should be stressed that for these exactly solvable systems, the inside of the square root 
$R_1\bigl(\mathcal{E}(n)\bigr)^2+4R_0\bigl(\mathcal{E}(n)\bigr)$ \eqref{alpm} is always 
a complete square and $\alpha_\pm\bigl(\mathcal{E}(n)\bigr)$ are polynomials in $n$ or $q^{\pm n}$
of maximal degree two. For the full details of these exactly solvable discrete quantum systems,
a paper by Odake and myself \cite{os12}
should be consulted.
\subsection{Krawtchouk and dual Hahn }
\label{sec:KrdHn}
These two systems have $\mathcal{E}(n)=n$ and $R_0=1$ and $R_1=0$ 
so that $\alpha_\pm\bigl(\mathcal{E}(n)\bigr)=\pm1$
and all the moments are identical
\begin{equation}
\text{K}, \text{dH}:\qquad \mu_{2m}=\mu_2=2\sum_{n=0}^{N-1}A_nC_{n+1}/\norm{\eta}^2, 
\quad m\in\mathbb{N}.
\label{mu2KdH}
\end{equation}
This means by \eqref{b123}
\begin{equation}
b_1^2=\mu_2,\quad b_2^2=1-\mu_2,\quad b_3^2=0.
\label{b30}
\end{equation}
These two examples shed light on an important property of the moments as stated in the following
\begin{prop}
\label{prop:scale}
When all the moments are equal, $\mu_{2m}=\mu_2$, $m\in\mathbb{N}$, 
the Lanczos orthogonalisation stops at $\mathcal{O}_2$. 
Likewise, when all the moments form a geometrical sequence $\mu_{2m}=\lambda^{2(m-1)}\mu_2$, 
$m\in\mathbb{N}$, $\lambda\in\mathbb{R}_{>0}$, 
the Lanczos orthogonalisation stops at $\mathcal{O}_2$. This case reduces to the constant case
by the scaling of the Hamiltonian $\mathcal{H}\to\lambda^{-1}\mathcal{H}$, which is
absorbed by the time rescaling $t\to\lambda t$.
The very early stopping of the Lanczos orthogonalisation may be considered as a clear sign of `non-complexity' of 
integrable systems.
\end{prop}
 The other data are
\paragraph{Krawtchouk}
\begin{align}
&B(x)=p(N-x),\quad D(x)=(1-p)x,\quad 0<p<1,\quad \eta(x)=x,\quad \mathcal{E}(n)=n,\n
& A_n=-p(N-n),\quad C_n=-(1-p)n,\quad \mu_2=2p(1-p)<1.
\label{Kmu2}
\end{align}
\paragraph{Dual Hahn} ($a,b>0$)
\begin{align}
&B(x)=\frac{(x+a)(x+a+b-1)(N-x)}
  {(2x-1+a+b)(2x+a+b)},
\quad
D(x)=\frac{x(x+b-1)(x+a+b+N-1)}
  {(2x-2+a+b)(2x-1+a+b)},\n
  &\eta(x)=x(x+a+b-1),\ \mathcal{E}(n)=n,\
  A_n=-(n+a)(N-n),\
  C_n=- n(b+N-n),\n
  &\mu_2=\frac{N+2}{10}\cdot\frac{4-5(a+b)+10ab-6N+5(a+b)N+2N^2}{2N+3(a+b)-2}.
  \label{dH1}
\end{align}
\subsection{affine $q$-Krawtchouk and dual $q$-Hahn}
\label{sec:qKrdHn}
These systems share the same functions $R_0$ and $R_1$
\begin{align*}
&R_0(\mathcal{H})=(q^{-\tfrac12}-q^{\tfrac12})^2(\mathcal{H}')^2,
\quad R_1(\mathcal{H})=(q^{-\tfrac12}-q^{\tfrac12})^2\mathcal{H}',\quad 
\mathcal{H}'\eqdef \mathcal{H}+1,\quad \mathcal{E}(n)=q^{-n}-1,\\
&R_1\bigl(\mathcal{E}(n)\bigr)^2+4R_0\bigl(\mathcal{E}(n)\bigr)=(q^{-1}-q)^2q^{-2n},\\
&\Longrightarrow \alpha_+\bigl(\mathcal{E}(n)\bigr)
=(q^{-1}-1)q^{-n},\quad \alpha_-\bigl(\mathcal{E}(n)\bigr)=(q-1)q^{-n}.
\end{align*}
For both systems $A_n$ and $C_n$ are quadratic polynomials in $q^n$ so that $\mu_{2m}$ can be 
exactly calculated in a closed form.
\paragraph{affine $q$-Krawtchouk} ($0<p<q^{-1}$)
\begin{align}
& B(x)=(q^{x-N}-1)(1-pq^{x+1}),\quad
  D(x)=pq^{x-N}(1-q^x),\quad
\mathcal{E}(n)=q^{-n}-1,\n
&  A_n=-(q^{n-N}-1)(1-pq^{n+1}),\quad
  C_n=-pq^{n-N}(1-q^n),\quad  \ \eta(x)=q^{-x}-1,\n
&\mu_{2m}=2p(q^{-1}-1)^{2m}\sum_{n=0}^{N-1}
q^{n+1-N}(q^{n-N}-1)(1-pq^{n+1}) (1-q^{n+1})q^{-2mn}/\norm{\eta}^2, 
\label{aqK1}\\
&\qquad \norm{\eta}^2=N+\frac{q^{-2N}\bigl(1-q^N)(1-q^N(1-2q)\bigr)}{1-q^2}.
\label{aqK2}
\end{align}
\paragraph{dual $q$-Hahn} ($0<a,b<1$)
\begin{align}
&B(x)=
  \frac{(q^{x-N}-1)(1-aq^x)(1-abq^{x-1})}
  {(1-abq^{2x-1})(1-abq^{2x})},\n
&D(x)=aq^{x-N-1}
  \frac{(1-q^x)(1-abq^{x+N-1})(1-bq^{x-1})}
  {(1-abq^{2x-2})(1-abq^{2x-1})},\n
&\mathcal{E}(n)=q^{-n}-1,\quad
  \eta(x)=(q^{-x}-1)(1-abq^{x-1}),\n
&A_n=-(1-aq^n)(q^{n-N}-1),\quad
  C_n =- aq^{-1}(1-q^n)(q^{n-N}-b),\n
&\mu_{2m}=2aq^{-1}(q^{-1}-1)^{2m}\sum_{n=0}^{N-1}
(1-aq^n)(q^{n-N}-1)(1-q^{n+1})(q^{n+1-N}-b)q^{-2mn}/\norm{\eta}^2.
\label{dqH1}
\end{align}
\subsection{quantum $q$-Krawtchouk }
\label{sec:qqKr}
The system has the functions $R_0$ and $R_1$
\begin{align*}
&R_0(\mathcal{H})=(q^{-\tfrac12}-q^{\tfrac12})^2(\mathcal{H}')^2,
\quad R_1(\mathcal{H})=(q^{-\tfrac12}-q^{\tfrac12})^2\mathcal{H}',\quad 
\mathcal{H}'\eqdef\mathcal{H}-1,\quad \mathcal{E}(n)=1-q^{n},\\
&R_1\bigl(\mathcal{E}(n)\bigr)^2+4R_0\bigl(\mathcal{E}(n)\bigr)=(q^{-1}-q)^2q^{2n},\\
&\Longrightarrow \alpha_+\bigl(\mathcal{E}(n)\bigr)=(1-q)q^{n},
\quad \alpha_-\bigl(\mathcal{E}(n)\bigr)=-(q^{-1}-1)q^{n}.
\end{align*}
Other data are ($p>q^{-N}$)
\begin{align*}
&B(x)=p^{-1}q^x(q^{x-N}-1),\quad
  D(x)=(1-q^x)(1-p^{-1}q^{x-N-1}),\
\mathcal{E}(n)=1-q^n,\\
&A_n=-p^{-1}q^{-n-N-1}(1-q^{N-n}),\quad
  C_n=-(q^{-n}-1)(1-p^{-1}q^{-n}),\quad
  \eta(x)=q^{-x}-1.
\end{align*}
The  $A_n$ and $C_n$ are quadratic polynomials in $q^{-n}$ so that $\mu_{2m}$ can be 
exactly calculated in a closed form,
\begin{align}
&\mu_{2m}=2p^{-1}(1-q)^{2m}\sum_{n=0}^{N-1}
q^{-n-N-1}(1-q^{N-n})(q^{-n-1}-1)(1-p^{-1}q^{-n-1})q^{2mn}/\norm{\eta}^2,
\label{qqKr}\\
&\qquad \norm{\eta}^2=N+\frac{q^{-2N}\bigl(1-q^N)(1-q^N(1-2q)\bigr)}{1-q^2}.\nonumber
\end{align}
\subsection{ $q$--Krawtchouk}
\label{sec:qKr}
The system has the functions $R_0$ and $R_1$ with $\mathcal{E}(n)=(q^{-n}-1)(1+pq^n)$ ($p>0$),
\begin{align*}
&R_0(\mathcal{H})=(q^{-\frac12}-q^{\frac12})^2
  \bigl(\mathcal{H}^{\prime\,2}+p(q^{-\frac12}+q^{\frac12})^2\bigr),\quad
R_1(\mathcal{H})=(q^{-\frac12}-q^{\frac12})^2 \mathcal{H}',\quad
  \mathcal{H}'\eqdef \mathcal{H}+1-p,\\
&R_1\bigl(\mathcal{E}(n)\bigr)^2+4R_0\bigl(\mathcal{E}(n)\bigr)=(q^{-1}-q)^2(q^{-n}+pq^n)^2,\\
&\Longrightarrow \alpha_+\bigl(\mathcal{E}(n)\bigr)=(q^{-1}-1)(q^{-n}+pq^{n+1})
\quad \alpha_-\bigl(\mathcal{E}(n)\bigr)=-(1-q)(q^{-n}+pq^{n-1}).
\end{align*}
Other data are ($p>0$)
\begin{align*}
&B(x)=q^{x-N}-1,\quad
  D(x)=p(1-q^x),\quad
\mathcal{E}(n)=(q^{-n}-1)(1+pq^n),\quad
  \eta(x)=q^{-x}-1,\\\
&A_n=-\frac{(q^{n-N}-1)(1+pq^n)}{(1+pq^{2n})(1+pq^{2n+1})}\,,
  \quad
  C_n=-pq^{2n-N-1}\frac{(1-q^n)(1+pq^{n+N})}
  {(1+pq^{2n-1})(1+pq^{2n})}.
\end{align*}
As  $A_n$ and $C_n$ are rational function of $q^{n}$, exact calculation of $\mu_{2m}$ is rather 
complicated,
\begin{align}
&\mu_{2m}=\frac{2p(q^{-1}-1)^{2m}}{\norm{\eta}^2}\sum_{n=0}^{N-1}
q^{2n-N+1}\frac{(q^{n-N}-1)(1+pq^n)}{(1+pq^{2n})(1+pq^{2n+1})}\n
&\hspace{45mm} \times \frac{(1-q^{n+1})(1+pq^{n+1+N})}
  {(1+pq^{2n+1})(1+pq^{2n+2})}(q^{-n}+pq^{n+1})^{2m},
\label{qKr}\\
&\qquad \norm{\eta}^2=N+\frac{q^{-2N}\bigl(1-q^N)(1-q^N(1-2q)\bigr)}{1-q^2}.\nonumber
\end{align}

\subsection{Hahn  and Racah}
\label{sec:HnRa}
These two systems have similar structures.
\paragraph{Hahn}
The system has the functions $R_0$ and $R_1$ with $\mathcal{E}(n)=n(n+a+b-1)$ ($a,b>0$),
\begin{align*}
&R_0(\mathcal{H})=4 \mathcal{H}+(a+b-2)(a+b),\quad 
  R_1=2,\\
&\Longrightarrow \alpha_+\bigl(\mathcal{E}(n)\bigr)=2n+a+b,\quad 
\alpha_-\bigl(\mathcal{E}(n)\bigr)=-(2n+a+b-2).
 \end{align*}
 Other data are 
\begin{align*}
&B(x)=(x+a)(N-x),\quad
  D(x)= x(b+N-x),\quad \eta(x)=x,\\
& A_n=-\frac{(n+a)(n+a+b-1)(N-n)}{(2n-1+a+b)(2n+a+b)},\quad
C_n=-\frac{n(n+b-1)(n+a+b+N-1)}{(2n-2+a+b)(2n-1+a+b)}.
\end{align*}
As  $A_n$ and $C_n$ are rational functions of $n$, exact calculation of $\mu_{2m}$ is rather 
complicated,
\begin{align}
&\mu_{2m}=\frac{2}{\norm{\eta}^2}\sum_{n=0}^{N-1}
\frac{(n+a)(n+a+b-1)(N-n)}{(2n-1+a+b)}\cdot\frac{(n+1)(n+b)(n+a+b+N)}{(2n+1+a+b)}\n
&\hspace{50mm} \times(2n+a+b)^{2(m-1)},
\label{Hn}\\
&\qquad \qquad \qquad \norm{\eta}^2=\frac{N(N+1)(2N+1)}6.\nonumber
\end{align}
\paragraph{Racah}
The system has the functions $R_0$ and $R_1$ with $\mathcal{E}(n)=n(n+\tilde{d})$ ($a,b>0$),
\begin{align*}
&R_0(\mathcal{H})=4 \mathcal{H}+\tilde{d}^2-1\quad 
  R_1=2,\quad \td\eqdef a+b-N-d-1,\quad \eta(x)=x(x+d),\\
&\Longrightarrow \alpha_+\bigl(\mathcal{E}(n)\bigr)=2n+\td+1,\quad 
\alpha_-\bigl(\mathcal{E}(n)\bigr)=-(2n+\td-1).
 \end{align*}
  Other data are ($d>0,\ a>N+d,\ 0<b<1+d$),
\begin{align*}
&B(x)
  =-\frac{(x+a)(x+b)(x-N)(x+d)}{(2x+d)(2x+d+1)},\quad
D(x)
  =-\frac{(x+d-a)(x+d-b)(x+d+N)x}{(2x+d-1)(2x+d)},\\
& A_n=\frac{(n+a)(n+b)(n-N)(n+\tilde{d})}
  {(2n+\tilde{d})(2n+\tilde{d}+1)},\quad
C_n= \frac{(n+\tilde{d}-a)(n+\tilde{d}-b)(n+\tilde{d}+N)n}
  {(2n+\tilde{d}-1)(2n+\tilde{d})}.
\end{align*}
As  $A_n$ and $C_n$ are rational functions of $n$, exact calculation of $\mu_{2m}$ is rather 
complicated,
\begin{align}
&\mu_{2m}=\frac{2}{\norm{\eta}^2}\sum_{n=0}^{N-1}
\frac{(n+a)(n+b)(n-N)(n+\tilde{d})}{(2n+\td)}\times(2n+\td+1)^{2(m-1)}\n
&\hspace{30mm}\times \frac{(n+1)(n+1+\td-a)(n+1+\td-b)(n+1+\tilde{d}+N)}{(2n+\td+2)},
\label{Ra}\\
&\qquad \qquad \qquad \norm{\eta}^2=\sum_{x=0}^Nx^2(x+d)^2.\nonumber
\end{align}

\subsection{$q$-Hahn  and $q$-Racah}
\label{sec:qHnqRa}
These two systems have similar structures.
\paragraph{$q$-Hahn}
The system has the functions $R_0$ and $R_1$ with $\mathcal{E}(n)=(q^{-n}-1)(1-abq^{n-1})$, 
\begin{align*}
 &R_0(\mathcal{H})=(q^{-\frac12}-q^{\frac12})^2
  \bigl(\mathcal{H}^{\prime\,2}-ab(1+q^{-1})^2\bigr),\ 
R_1(\mathcal{H})=(q^{-\frac12}-q^{\frac12})^2\mathcal{H}',\ 
 \mathcal{H}'\eqdef \mathcal{H}+1+abq^{-1},\\
&R_1\bigl(\mathcal{E}(n)\bigr)^2+4R_0\bigl(\mathcal{E}(n)\bigr)
=(q^{-1}-q)^2(q^{-n}-abq^{n-1})^2,\\
&\Longrightarrow \alpha_+\bigl(\mathcal{E}(n)\bigr)=(q^{-1}-1)(q^{-n}-abq^n),\quad 
\alpha_-\bigl(\mathcal{E}(n)\bigr)=-(1-q)(q^{-n}-abq^{n-2}).
 \end{align*}
Other data are ($0<a,b<1$),
\begin{align*}
&B(x) =(1-aq^x)(q^{x-N}-1),\quad
  D(x)= aq^{-1}(1-q^x)(q^{x-N}-b),\quad  \eta(x)=q^{-x}-1,\\
& A_n=-\frac{(q^{n-N}-1)(1-aq^n)(1-abq^{n-1})}{(1-abq^{2n-1})(1-abq^{2n})},\\
&C_n=-aq^{n-N-1}\,
  \frac{(1-q^n)(1-abq^{n+N-1})(1-bq^{n-1})}{(1-abq^{2n-2})(1-abq^{2n-1})}.
\end{align*}
As  $A_n$ and $C_n$ are rational function of $q^n$, exact calculation of $\mu_{2m}$ is rather 
complicated,
\begin{align}
&\mu_{2m}=\frac{2a(q^{-1}-1)^{2m}}{\norm{\eta}^2}\sum_{n=0}^{N-1}q^{-2mn+n-N}
\frac{(q^{n-N}-1)(1-aq^n)(1-abq^{n-1})}{(1-abq^{2n-1})}\cdot(1-abq^{2n})^{2(m-1)}\n
&\hspace{45mm}\times  \frac{(1-q^{n+1})(1-abq^{n+N})(1-bq^{n})}{(1-abq^{2n+1})},
\label{qHn}\\
&\qquad \qquad \qquad \norm{\eta}^2=\sum_{x=0}^N(q^{-x}-1)^2.\nonumber
\end{align}
\paragraph{$q$-Racah}
The system has the functions $R_0$ and $R_1$ with $\mathcal{E}(n)=(q^{-n}-1)(1-\td q^{n})$, 
\begin{align*}
 &R_0(\mathcal{H})=(q^{-\frac12}-q^{\frac12})^2
  \bigl(\mathcal{H}^{\prime\,2}-(q^{-\frac12}+q^{\frac12})^2\td\bigr),\ 
R_1(\mathcal{H})=(q^{-\frac12}-q^{\frac12})^2\mathcal{H}',\ 
 \mathcal{H}'\eqdef \mathcal{H}+1+\td,\\
&R_1\bigl(\mathcal{E}(n)\bigr)^2+4R_0\bigl(\mathcal{E}(n)\bigr)
=(q^{-1}-q)^2(q^{-n}-\td q^{n})^2,\\
&\Longrightarrow \alpha_+\bigl(\mathcal{E}(n)\bigr)=(q^{-1}-1)(q^{-n}-\td q^{n+1}),\quad 
\alpha_-\bigl(\mathcal{E}(n)\bigr)=-(1-q)(q^{-n}-\td q^{n-1}).
 \end{align*}
Other data are ($\tilde{d}\eqdef abd^{-1}q^{-N-1}$, $0<d<1,\ 0<a<q^Nd,\ qd<b<1$),
\begin{align*}
&B(x) =-\frac{(1-aq^x)(1-bq^x)(1-q^{x-N})(1-dq^x)}
  {(1-dq^{2x})(1-dq^{2x+1})},\quad \eta(x)=(q^{-x}-1)(1-dq^x),\\
&  D(x)= -\tilde{d}\,
  \frac{(1-a^{-1}dq^x)(1-b^{-1}dq^x)(1-dq^{N+x})(1-q^x)}
  {(1-dq^{2x-1})(1-dq^{2x})},\\
& A_n=\,\frac{(1-aq^n)(1-bq^n)(1-q^{n-N})(1-\tilde{d}q^n)}
  {(1-\tilde{d}q^{2n})(1-\tilde{d}q^{2n+1})},\\
&C_n=d\,
  \frac{(1-a^{-1}\tilde{d}q^n)(1-b^{-1}\tilde{d}q^n)(1-\tilde{d}q^{n+N})
  (1-q^n)}
  {(1-\tilde{d}q^{2n-1})(1-\tilde{d}q^{2n})}.
\end{align*}
As  $A_n$ and $C_n$ are rational function of $q^n$, exact calculation of $\mu_{2m}$ is rather 
complicated,
\begin{align}
&\mu_{2m}=\frac{2d(q^{-1}-1)^{2m}}{\norm{\eta}^2}\!\sum_{n=0}^{N-1}q^{-2mn}
\frac{(1-aq^n)(1-bq^n)(1-q^{n-N})(1-\tilde{d}q^n)}
  {(1-\tilde{d}q^{2n})}\cdot(1-\td q^{2n+1})^{2(m-1)}\n
&\hspace{45mm}\times  
\frac{(1-a^{-1}\tilde{d}q^{n+1})(1-b^{-1}\tilde{d}q^{n+1})(1-\tilde{d}q^{n+1+N})
  (1-q^{n+1})}
  {(1-\tilde{d}q^{2n+2})},
\label{qRa}\\
&\qquad \qquad \qquad \norm{\eta}^2=\sum_{x=0}^N(q^{-x}-1)^2(1-dq^x)^2.\nonumber
\end{align}

\section{Evaluation of $\mu_m$ of other exactly solvable quantum mechanical systems}
 \label{sec:mumother}
In order to evaluate the moments $\{\mu_m\}$ for verifications of
Krylov complexity of general quantum mechanical systems through
orthonormalisation of a Krylov subspace, an appropriate definition of the inner product of
operators is essential. The simplest trace one 
$(\mathcal{V},\mathcal{W})=\text{Tr}\lbr\mathcal{V}^\dagger\mathcal{W}\rbr$ \eqref{indef}
is obviously ill-defined for most operators, which are unbounded.
A simple prescription to deal with unbounded operators is to introduce the finite temperature (T) effects through
the Boltzmann factor $e^{-\beta \mathcal{H}}$, $\beta=1/T$. 
The finite temperature inner product suppresses the contributions of 
higher energy eigenstates and make the trace summable. 
The general forms of the inner product including the Boltzmann factor were discussed in detail in section VIII
of \cite{parker} and \cite{caputa}.
Here I choose, following \cite{parker,caputa}, the Wightman inner product
\begin{equation}
(\mathcal{V},\mathcal{W})_\beta\eqdef
\frac1{Z}\text{Tr}\lbr e^{-\beta\mathcal{H}/2}\mathcal{V}^\dagger
e^{-\beta\mathcal{H}/2}\mathcal{W}\rbr
=\frac1{Z}\sum_{n=0}^\infty\langle n\left|e^{-\beta\mathcal{H}/2}\mathcal{V}^\dagger
e^{-\beta\mathcal{H}/2}\mathcal{W}\right|n\rangle,
\label{binn}
\end{equation}
in which
\begin{equation*}
Z\eqdef\text{Tr}\lbr e^{-\beta\mathcal{H}}\rbr,\quad \beta=1/T.
\end{equation*}
However, as shown later in the examples \S\ref{sec:MeiChar}--\S\ref{sec:GegJac}, the factor $Z$ 
cancels out in the calculation of the moments.
The norm of an operator is defined by
\begin{equation}
(\mathcal{V},\mathcal{W})_\beta=(\mathcal{W}^\dagger,\mathcal{V}^\dagger)_\beta=(\mathcal{W},\mathcal{V})_\beta^*
\rightarrow \norm{\mathcal{V}}_\beta^2\eqdef (\mathcal{V},\mathcal{V})_\beta,
\label{bnorm}
\end{equation}
With this {\em symmetric} inner product the flip property 
$(\mathcal{V},\mathcal{L}\mathcal{W})_\beta
=(\mathcal{L}\mathcal{V},\mathcal{W})_\beta$ \eqref{Lflip}
holds 
and $\mathcal{V}^\dagger=\pm\mathcal{V}$  means 
\begin{equation}
(\mathcal{V},\mathcal{L}\mathcal{V})_\beta=\bigl((\mathcal{L}\mathcal{V})^\dagger,\mathcal{V}^\dagger)_\beta
=(-\mathcal{L}\mathcal{V},\mathcal{V})_\beta=-(\mathcal{V},\mathcal{L}\mathcal{V})_\beta=0.
\label{symmglip}
\end{equation}
Thus the  Lanczos orthonormalisation reviewed in section
\ref{sec:set} works equally  for the Wightman  inner product \eqref{binn}.

It should be stressed that the situation is different for {\em non-symmetric} inner products,
for example \cite{dymarsky},
\begin{equation*}
(\mathcal{V},\mathcal{W})_\rho\eqdef\frac1{Z}\text{Tr}\lbr \mathcal{V}^\dagger\rho_1\mathcal{W}\rho_2\rbr
=(\mathcal{W},\mathcal{V})_\rho^*\neq(\mathcal{W}^\dagger,\mathcal{V}^\dagger)_\rho,
\end{equation*}
in which $\rho_1,\rho_2$ are positive functions of the Boltzmann factor.
For $\rho_1\neq\rho_2$,  the flip property $(\mathcal{V},\mathcal{L}\mathcal{W})_\beta
=(\mathcal{L}\mathcal{V},\mathcal{W})_\beta$ \eqref{Lflip}
holds  but $\mathcal{V}^\dagger=\pm\mathcal{V}$ does not mean 
$(\mathcal{V},\mathcal{L}\mathcal{V})_\beta=0$. 
Therefore the  Lanczos orthonormalisation must be modified,   as shown in \cite{dymarsky}.

\bigskip
The Hamiltonian $\mathcal{H}$ and the sinusoidal coordinate $\eta$  of the 
exactly solvable quantum mechanical systems demonstrated in \cite{os8,os7,os12,os13} 
all satisfy the same relationship
\begin{equation*}
  [\mathcal{H},[\mathcal{H},\eta]\,]=\eta\,R_0(\mathcal{H})
  +[\mathcal{H},\eta]\,R_1(\mathcal{H})+R_{-1}(\mathcal{H}),
  \tag{\ref{twocom}}
\end{equation*}
therefore the essential formulas
\begin{align*}
  e^{it\mathcal{H}}\eta\,e^{-it\mathcal{H}}
  &=\mathcal{L}\eta\cdot
  \frac{e^{i\alpha_+(\mathcal{H})t}-e^{i\alpha_-(\mathcal{H})t}}
  {\alpha_+(\mathcal{H})-\alpha_-(\mathcal{H})}
  -R_{-1}(\mathcal{H})/R_{0}(\mathcal{H})\n
  &\quad
  +\bigl(\eta+R_{-1}(\mathcal{H})/R_0(\mathcal{H})\bigr)
  \frac{-\alpha_-(\mathcal{H})e^{i\alpha_+(\mathcal{H})t}
  +\alpha_+(\mathcal{H})e^{i\alpha_-(\mathcal{H})t}}
  {\alpha_+(\mathcal{H})-\alpha_-(\mathcal{H})}.
  \tag{\ref{quantsol}}
\end{align*}
and
\begin{align*}
 \mathcal{L}^m\eta=&\,\eta\,\mathcal{A}_m(\mathcal{H})
  + \mathcal{L}\eta\,\mathcal{B}_m(\mathcal{H})
  +\mathcal{C}_m(\mathcal{H}),\qquad m\in\mathbb{Z}_{\ge0},
\tag{\ref{Lmcom}}
\end{align*}
for the evaluation of the moments $\{\mu_m\}$
hold with the same coefficients $\mathcal{A}_m(\mathcal{H})$ \eqref{amf},
$\mathcal{B}_m(\mathcal{H})$ \eqref{bmf} and $\mathcal{C}_m(\mathcal{H})$ \eqref{cmf}.

Let us evaluate the moments
\begin{align}
\mu_m=(\mathcal{O}_0,\mathcal{L}^m\mathcal{O}_0)_\beta
\eqdef\frac1{Z\norm{\eta}_\beta^2}\text{Tr}\lbr e^{-\beta\mathcal{H}/2}\eta
e^{-\beta\mathcal{H}/2}\mathcal{L}^m\eta\rbr,\qquad m\in\mathbb{Z}_{\ge0}.
\end{align}
Since the denominator $Z\norm{\eta}_\beta^2$ is independent of $m$, 
the numerator only
\begin{equation}
\tilde{\mu}_m\eqdef\text{Tr}\lbr e^{-\beta\mathcal{H}/2}\eta
e^{-\beta\mathcal{H}/2}\mathcal{L}^m\eta\rbr
=\sum_{n=0}^\infty\langle n\left|e^{-\beta\mathcal{H}/2}\eta
e^{-\beta\mathcal{H}/2}\mathcal{L}^m\eta\right|n\rangle
\label{tilmudef}
\end{equation}
will be discussed henceforth.

The orthonormal eigenvectors of the Hamiltonian $\mathcal{H}$,  $\{|n\rangle\}$ 
have the same structure as \eqref{ndef1} and \eqref{ndef2}. 
But the forms of the sinusoidal coordinates $\eta(x)$ and 
the energy spectrum $\mathcal{E}(n)$ are not restricted to those listed in 
\eqref{5eta} and \eqref{En}.
As the types of the eigenpolynomials $\{P_n(\eta)\}$ are more varied than those
of the discrete quantum mechanics,  the uniform normalisation condition 
$P_n(0)=1$ \eqref{unicon} no longer applies and the three term recurrence
relation has a more general form
\begin{align}
  \eta P_n(\eta)&=A_nP_{n+1}(\eta)+B_nP_{n}(\eta)
  +C_nP_{n-1}(\eta).
  \label{3termb}
\end{align}
This simply means that $A_n+C_n$ in the simplification of the moments in section 
\ref{sec:mu2m} should be replaced by $-B_n$.

\bigskip
The simplification of $\langle n\left|e^{-\beta\mathcal{H}/2}\eta
e^{-\beta\mathcal{H}/2}\mathcal{L}^m\eta\right|n\rangle$ goes in parallel 
with that in section \ref{sec:mu2m} as follows
\begin{align}
&\langle n\left|e^{-\beta\mathcal{H}/2}\eta
e^{-\beta\mathcal{H}/2}\mathcal{L}^m\eta\right|n\rangle
=\sum_{l=n,n\pm1}e^{-\beta(\mathcal{E}(n)+\mathcal{E}(l))/2}\langle n|\eta|l\rangle
\langle l|\mathcal{L}^m\eta|n\rangle.
\end{align}
The three parts (a) $l=n$, (b) $l=n+1$ and (c) $l=n-1$ ($n\ge1$) are separately simplified.
\paragraph{(a) $l=n$ part}
This part vanishes as shown in section \ref{sec:mu2m}
\begin{align*}
&\langle n|\eta|n\rangle
\langle n|\mathcal{L}^m\eta|n\rangle=
\langle n|\eta|n\rangle\left\{\langle n|\eta|n\rangle\mathcal{A}_m\bigl(\mathcal{E}(n)\bigr)+
\langle n|\mathcal{L}\eta|n\rangle\mathcal{B}_m\bigl(\mathcal{E}(n)\bigr)
+\mathcal{C}_m\bigl(\mathcal{E}(n)\bigr)\right\}=0,
\end{align*}
as $\langle n|\eta|n\rangle^2\mathcal{A}_m\bigl(\mathcal{E}(n)\bigr)$ cancels 
$\langle n|\eta|n\rangle\mathcal{C}_m\bigl(\mathcal{E}(n)\bigr)$ part and 
$\langle n|\mathcal{L}\eta|n\rangle=0$.
\paragraph{(b) $l=n+1$ part}
This part is
\begin{align}
&\langle n|\eta|n+1\rangle
\langle n+1|\mathcal{L}^m\eta|n\rangle\n
&=\langle n|\eta|n+1\rangle\left\{\langle n+1|\eta|n\rangle\mathcal{A}_m\bigl(\mathcal{E}(n)\bigr)+
\langle n+1|\mathcal{L}\eta|n\rangle\mathcal{B}_m\bigl(\mathcal{E}(n)\bigr)
\right\}\n
&=A_nC_{n+1}\left\{\mathcal{A}_{m}\bigl(\mathcal{E}(n)\bigr)
+\alpha_+\bigl(\mathcal{E}(n)\bigr)\mathcal{B}_{m}\bigl(\mathcal{E}(n)\bigr)\right\}\n
&=A_nC_{n+1}\alpha_+\bigl(\mathcal{E}(n)\bigr)^{m},
\end{align}
as in \eqref{plussum}.
\paragraph{(b) $l=n-1$ part}
This part is
\begin{align}
&\langle n|\eta|n-1\rangle
\langle n-1|\mathcal{L}^m\eta|n\rangle\n
&=\langle n|\eta|n-1\rangle\left\{\langle n-1|\eta|n\rangle\mathcal{A}_m\bigl(\mathcal{E}(n)\bigr)+
\langle n-1|\mathcal{L}\eta|n\rangle\mathcal{B}_m\bigl(\mathcal{E}(n)\bigr)
\right\}\n
&=A_{n-1}C_{n}\left\{\mathcal{A}_{m}\bigl(\mathcal{E}(n)\bigr)
+\alpha_-\bigl(\mathcal{E}(n)\bigr)\mathcal{B}_{m}\bigl(\mathcal{E}(n)\bigr)\right\}\n
&=A_{n-1}C_{n}\alpha_-\bigl(\mathcal{E}(n)\bigr)^{m},\quad n\ge1,
\end{align}
as in \eqref{minsum}.
These lead to
\begin{align}
&\langle n\left|e^{-\beta\mathcal{H}/2}\eta
e^{-\beta\mathcal{H}/2}\mathcal{L}^m\eta\right|n\rangle\n
&=e^{-\beta(\mathcal{E}(n)+\mathcal{E}(n+1))/2}A_nC_{n+1}\alpha_+\bigl(\mathcal{E}(n)\bigr)^{m}
 +e^{-\beta(\mathcal{E}(n)+\mathcal{E}(n-1))/2}A_{n-1}C_{n}\alpha_-\bigl(\mathcal{E}(n)\bigr)^{m},
\label{nbetaform}
\end{align}
and
\begin{align}
\tilde{\mu}_m
&=\sum_{n=0}^\infty\langle n\left|e^{-\beta\mathcal{H}/2}\eta
e^{-\beta\mathcal{H}/2}\mathcal{L}^m\eta\right|n\rangle\n
&=2\sum_{n=0}^\infty
e^{-\beta(\mathcal{E}(n)+\mathcal{E}(n+1))/2}A_nC_{n+1}
\alpha_+\bigl(\mathcal{E}(n)\bigr)^{m}
\times
\left\{
\begin{array}{cr}
1  &   m:\ \text{even}   \\
0  &   m:\ \text{odd}   
\end{array}
\right.
\label{finmumb}
\end{align}
The squared norm of the sinusoidal coordinate $\eta$ is simplified as before,
\begin{align}
Z\norm{\eta}_\beta^2&=\text{Tr}\lbr e^{-\beta\mathcal{H}/2}\eta e^{-\beta\mathcal{H}/2}\eta\rbr\n
&=\sum_{n=0}^\infty\sum_{l=n,n\pm1}e^{-\beta\mathcal{E}(n)/2}e^{-\beta\mathcal{E}(l)/2}
\langle n|\eta|l\rangle\langle l|\eta|n\rangle\n
&=\sum_{n=0}^\infty\left\{
e^{-\beta\mathcal{E}(n)}\langle n|\eta|n\rangle\langle n|\eta|n\rangle
+e^{-\beta\bigl(\mathcal{E}(n)+\mathcal{E}(n+1)\bigr)/2}
\langle n|\eta|n+1\rangle\langle n+1|\eta|n\rangle\right.\n
&\hspace{50mm} \left. 
+e^{-\beta\bigl(\mathcal{E}(n)+\mathcal{E}(n-1)\bigr)/2}
\langle n|\eta|n-1\rangle\langle n-1|\eta|n\rangle\right\}\n
&=\sum_{n=0}^\infty\left\{
e^{-\beta\mathcal{E}(n)}B_n^2+2e^{-\beta\bigl(\mathcal{E}(n)+\mathcal{E}(n+1)\bigr)/2}A_nC_{n+1}
\right\}.
\end{align}
These results are summarised as  the following
\begin{theo}
\label{theo:mainb}
The moments of the exactly solvable ordinary quantum systems have a very simple exact expression
\begin{equation}
\mu_{2m}=\frac2{Z\norm{\eta}_\beta^2}
\sum_{n=0}^{\infty}e^{-\beta(\mathcal{E}(n)+\mathcal{E}(n+1))/2}
A_nC_{n+1}\left(\alpha_+\bigl(\mathcal{E}(n)\bigr)\right)^{2m},
\quad \mu_{2m-1}=0,\quad m\in\mathbb{N}.
\label{mainthb}
\end{equation}
It applies to two non-compact discrete quantum systems with the eigenpolynomials, the  Meixner (M) and 
Charlier (C), as well as the exactly solvable systems related to the Hermite (H), Laguerre (L), Gegenbauer (G),
and Jacobi (J)  polynomials.
\end{theo}
In the rest of this section, the necessary data for the evaluation of the moments $\mu_{2m}$ of 
the above six exactly solvable quantum mechanical systems are provided.
For the ordinary one-dimensional quantum systems, they are the potential $U(x)$ in the  Hamiltonian 
$\mathcal{H}=p^2+U(x)$ in which $p$ is the canonical momentum operator conjugate to $x$, $[p,x]=-i$, 
the sinusoidal coordinates $\eta(x)$, the energy eigenvalues $\mathcal{E}(n)$, $R_0(\mathcal{H})$ and
$R_1(\mathcal{H})$ for the  derivation of $\alpha_\pm\bigl(\mathcal{E}(n)\bigr)$ and the coefficients of the
three term recurrence relation $A_n$ $B_n$ and $C_n$.

\subsection{Meixner and Charlier}
\label{sec:MeiChar}
These two are non-compact exactly solvable discrete quantum systems. 
For the summability of the trace, the finite temperature effects \eqref{binn} are needed.
These two systems  have very simple structure
\begin{align*}
M, C: \quad &\mathcal{E}(n)=n,\quad \eta(x)=x,\\
&R_0=1,\quad R_1=0\ \Rightarrow \ \alpha_+\bigl(\mathcal{E}(n)\bigr)=1,\quad 
\alpha_-\bigl(\mathcal{E}(n)\bigr)=-1.
\end{align*}
This simply means that all the moments are equal $\mu_{2m}=\mu_2$, $m\in\mathbb{N}$.
The Lanczos  orthonormalisation stops at $\mathcal{O}_2$.
The other data are
\begin{align}
M: \quad & B(x)=\frac{c}{1-c}(x+b),\quad
  D(x)=\frac{x}{1-c},\quad 0<c<1,\quad 0<b,\n
  \quad & A_n=-\frac{c}{1-c}(n+b),\quad
  C_n=-\frac{n}{1-c},\n
  &\mu_2=\frac{2c}{(1-c)^2Z\norm{\eta}_\beta^2}
  \sum_{n=0}^\infty e^{-\beta(2n+1)/2}(n+1)(n+b),
  \label{Mmu}\\
C: \quad &  B(x)=a,\quad D(x)=x,\quad A_n=-a,\quad C_n=-n, \quad 0<a,\n
&\mu_2=\frac{2a}{Z\norm{\eta}_\beta^2} \sum_{n=0}^\infty e^{-\beta(2n+1)/2}(n+1).
\label{Cmu}
\end{align}
As expected, the Boltzmann factor $e^{-\beta\mathcal{H}}$
 introduced for the trace definition in \ref{binn} makes 
the infinite sum of these two $\mu_2$'s convergent. The summation can be carried out explicitly.
\subsection{Hermite and Laguerre }
\label{sec:HerLag}
These two are the best known examples of exactly solvable one-dimensional quantum mechanical systems,
the harmonic oscillator  for the Hermite (H) and the harmonic oscillator with a centrifugal potential
for the Laguerre (L).
Like K and dH cases in \S\ref{sec:KrdHn}, all the moments form geometrical sequences,
$\mu_{2m}=\lambda^{2(m-1)}\mu_2$, $\lambda=2$ for H and $\lambda=4$ for L.
The Krylov orthogonalisation stops at $\mathcal{O}_2$, a clear sign of `non-complexity' of these solvable 
systems.
The data for each system are:
\paragraph{Hermite}
\begin{align}
& U(x)=x^2-1,\quad -\infty<x<\infty,\quad \mathcal{E}(n)=2n,\quad \eta(x)=x,\n
& R_0=4,\quad R_1=0,\quad \alpha_+\bigl(\mathcal{E}(n)\bigr)=2,
\quad \alpha_-\bigl(\mathcal{E}(n)\bigr)=-2,\n
&  A_n=\frac12,\quad B_n=0,\quad C_n=n,\n
& \Longrightarrow \mu_{2m}=2^{2m}.
\label{Hmum}
\end{align}
This is a very well-known result.
\paragraph{Laguerre} ($g>1$)
\begin{align}
 & U(x)=x^2+g(g-1)/x^2-(1+2g),\quad  0<x<\infty,\quad \mathcal{E}(n)=4n,\quad \eta(x)=x^2,\n
& R_0=16,\quad R_1=0,\quad \alpha_+\bigl(\mathcal{E}(n)\bigr)=4,
\quad \alpha_-\bigl(\mathcal{E}(n)\bigr)=-4,\n
&A_n=-(n+1), \quad B_n=2n+g+1/2,\quad C_n=-(n+g-1/2),\n
&\mu_{2m}=\frac{2^{4m+1}}{Z\norm{\eta}_\beta^2}
\sum_{n=0}^{\infty}e^{-2\beta(2n+1)}(n+1)(n+g+1/2).
\label{Lmu}
\end{align}
As mentioned above, the Heisenberg operator solution for $x^2+1/x^2$ (L) and $1/\sin^2x$ (G) 
potentials are reported by Nieto and Simmons \cite{nieto1}--\cite{nieto4}.
\subsection{Gegenbauer and Jacobi}
\label{sec:GegJac}
These two systems seem to  provide very interesting materials for verifying Krylov complexity.
They are defined on a finite line segment, $0<x<\pi$ for G and $0<x<\pi/2$ for J.
They both have quadratic energy spectrum in $n$.
The expression of $\mu_{2m}$ \eqref{mainthb} in the {\bf Theorem \ref{theo:mainb}} applies.
Here are their data:
\paragraph{Gegenbauer} ($g>1$)
\begin{align}
& U(x)=\frac{g(g-1)}{\sin^2x}-g^2,\quad 0<x<\pi,
\quad \mathcal{E}(n)=n(n+2g),\quad \eta(x)=\cos x,\n
&R_0(\mathcal{H})=4\mathcal{H}'-1,\quad \mathcal{H}'\eqdef \mathcal{H}+g^2,\quad R_1=2,
\quad \alpha_\pm(\mathcal{H})=1\pm2\sqrt{\mathcal{H}'},\n
&\alpha_+\bigl(\mathcal{E}(n)\bigr)=2(n+g)+1,
\quad \alpha_-\bigl(\mathcal{E}(n)\bigr)=-(2n+2g-1),\n
& A_n=\frac{n+1}{2(n+g)},\quad B_n=0,\quad C_n=\frac{n+2g-1}{2(n+g)}.\nonumber
\end{align}
\paragraph{Jacobi} ($g,h>1$)
\begin{align*}
& U(x)=\frac{g(g-1)}{\sin^2x}+\frac{h(h-1)}{\cos^2x}-(g+h)^2,
\quad  0<x<\pi/2,\\
& \mathcal{E}(n)=4n(n+g+h),\quad  \eta(x)=\cos 2x,\\
&R_0(\mathcal{H})=16(\mathcal{H}'-1),\quad\mathcal{H}'\eqdef\mathcal{H}+(g+h)^2,\quad R_1=8,
\quad \alpha_\pm(\mathcal{H})=4\pm4\sqrt{\mathcal{H}'},\\
&\alpha_+\bigl(\mathcal{E}(n)\bigr)=4(2n+g+h+1),
\quad \alpha_-\bigl(\mathcal{E}(n)\bigr)=-4(2n+g+h-1),\\
& A_n=\frac{2(n+1)(n+g+h)}{(2n+g+h)(2n+g+h+1)},\quad
B_n=\frac{(h-g)(g+h-1)}{(2n+g+h-1)(2n+g+h+1)},\\
& C_n=\frac{2(n+g-1/2)(n+h-1/2)}{(2n+g+h-1)(2n+g+h)}.
\end{align*}
\section{Summary and comments}
\label{sec:comments}
For a group of exactly solvable compact discrete quantum systems, the moments of the
operators in a Krylov subspace spanned by a Liouville operator $\mathcal{L}:=[\mathcal{H},\cdot]$
and the sinusoidal coordinate $\eta$ are evaluated explicitly.
They provide the essential tool for measuring the growth of operators evolving under Hamiltonian
dynamics \cite{parker}.
Understanding of the complexity of exactly solvable quantum systems
would reveal the nature of quantum chaos by contrast.
The moments of exactly solvable {\em non-compact} discrete dynamics and ordinary one-dimensional
quantum systems are also evaluated explicitly by adopting Wightman inner product involving the Boltzmann factor.
These exactly solvable discrete quantum mechanical systems 
can be regarded as a very special type of matrix models.

There are many exactly solvable ordinary quantum mechanical systems having finitely many discrete energy levels.
Among them, for example, the Morse potential and the soliton potential ($-1/\cosh^2x$) are
exactly solvable in the Heisenberg picture, too \cite{os7} and the formula 
$\mathcal{L}^m\eta=\,\eta\,\mathcal{A}_m(\mathcal{H})
  + \mathcal{L}\eta\,\mathcal{B}_m(\mathcal{H})+\mathcal{C}_m(\mathcal{H})$ \eqref{Lmcom} is 
  applicable.
  It is quite natural to expect that their `complexity' is qualitatively different from that of the systems
  having infinitely many energy levels only.
  I cannot apply the Lanczos algorithm to such systems as I do not know how to include the contribution of 
  the continuous energy levels in the operator inner product.

Another type of exactly solvable discrete quantum mechanical systems is also known \cite{os7,os13}
and they are also solvable in the Heisenberg picture.
Their eigenvectors contain the hypergeometric orthogonal polynomials of Askey scheme,
the Wilson, Askey-Wilson, continuous (dual) ($q$) Hahn, Meixner-Pollaczek, Al-Salam-Chihara,
continuous (big) $q$-Hermite, continuous $q$-Jacobi (Laguerre) polynomials.
It is expected that the moments of these quantum systems can be evaluated explicitly in a similar manner.

Exact Heisenberg operator solutions are also known for a family of multi-particle dynamics,
the Calogero models based on any root systems \cite{os9}.
It is a good challenge to generalise the present method for multi-particle systems.

Four explicit examples of multivariate discrete orthogonal polynomials, the multivariate Krawtchouk, Meixner 
and two types of Rahman polynomials, are constructed recently by myself \cite{mpoly1}--\cite{mpoly3}.
They are eigenvectors of respective Hamiltonians having nearest neighbour and other types of interactions.
Investigation of their complexity through Krylov orthonormalisation 
would expose the contrast between integrable and chaotic multi-particle dynamics.

One of the motivations of the present research is to display the main ingredients of Krylov complexity 
with the explicit description of the Hamiltonian $\mathcal{H}$ and the operator $\eta$, 
which is lacking in some reports.
%
%
\section*{Acknowledgements}
\label{sec:ack}
R.\,S. thanks Jen-Chi Lee for the hospitality and vigorous discussions during his visit to
National Yang Ming Chiao Tung University.
He also thanks Yi Yang for opening his eyes to the topic of Krylov complexity.
Many useful suggestions by referees are greatly acknowledged. 

\section*{Declarations}
\begin{itemize}
\item Funding: No funds, grants, or other support was received.
\item Data availability statement: Data sharing not applicable to this article as 
no datasets were generated or analysed during the current study.
\item Competing Interests: The author has no competing interests to declare that 
are relevant to the content of this article.
\end{itemize}



\begin{thebibliography}{99}

 
 \bibitem{parker}
D. E. Parker, X. Cao, A. Avdoshkin, T. Scaffidi and E. Altman, ``A Universal Operator
Growth Hypothesis," Phys. Rev. {\bf X9} (2019) 041017, 
{\tt arXiv:1812.08657v5\hspace{0pt}[cond-mat.stat-mech]}.


\bibitem{barbon}
J.\, L.\, Barb\'on, E.\, Rabinovici, R.\, Shir and R.\, Sinha, ``On The Evolution
  Of Operator Complexity Beyond Scrambling,"
  JHEP {\bf 10}
  (2019) 264, {\tt arXiv:1907.05393[hep-th]}.
  
 \bibitem{dymarsky}
 A.\, Dymarsky and A.\, Gorsky, ``Quantum chaos as delocalization in Krylov space,"
 Physical Review {\bf B 102}  (2020) 085137, {\tt arXov:1912.12227[cond-mat]}.
 
\bibitem{rabino}
E.\, Rabinovici, A.\, S\'anchez-Garrido, R.\, Shir and J.\, Sonner, ``Operator
  complexity: a journey to the edge of Krylov space,"
 JHEP {\bf 06}
  (2021)  062, {\tt arXiv:2009.01862[hep-th]}.
 
 
\bibitem{caputa}
P.\, Caputa, J.\, M.\, Magan and D.\, Patramanis,
``Geometry of Krylov Complexity," Phys. Rev. Research {\bf 4} (2022) 013041,
{\tt arXiv:2109.03824v2[hep-th]}.

\bibitem{yyang}
W.\,M\"uck and Y.\, Yang,
``Krylov complexity and orthogonal polynomials,"
Nucl. Phys. {\bf B984} (2022) 115948, {\tt arXiv:2205.12815[hep-th]}.

\bibitem{pratik}
B.\, Bhattacharjee, X.\, Cao, P.\, Nandy and T.\, Pathak,
``Operator growth in open quantum systems: lessons from the dissipative SYK,"
JHEP 03 (2023) 054, 
{\tt arXiv:2212.06180[quant-ph]}.

\bibitem{sachdev}
 S.\, Sachdev and J. Ye, ``Gapless spin fluid ground state in a random, 
 quantum Heisenberg ferromagnet,"  Phys. Rev. Lett. {\bf 70} (1993) 3339, 
 {\tt arXiv:cond-mat/9212030}.
 
\bibitem{kitaev}
A. Kitaev, ``A simple model of quantum holography," (2015).

\bibitem{roberts}
D.\, Roberts, D.\, Stanford and A.\, Streicher,
``Operator growth in the SKY model,"
JHEP {\bf 2018} 122 (2018).

\bibitem{askey}
G.\,E.\, Andrews, R.\, Askey  and R.\, Roy,
{\it Special Functions},
Encyclopedia of mathematics and its applications,
Cambridge Univ. Press, Cambridge, (1999).


\bibitem{ismail}
M.\,E.\,H.\, Ismail,
{\it Classical and quantum orthogonal polynomials in one variable\/},
Encyclopedia of mathematics and its applications, Cambridge Univ. Press, Cambridge, (2005).

\bibitem{koeswart} 
R.\, Koekoek,  P.\,A.\, Lesky  and R.\,F.\, Swarttouw,
{\it Hypergeometric orthogonal polynomials and their $q$-analogues,\/}
Springer Monographs in Mathematics, 
Springer-Verlag, Berlin, (2010).
%
\bibitem{krylov}
A.\,N. Krylov, ``On the numerical solution of the equation by which in technical questions 
frequencies of small oscillations of material systems are determined,"  
Izvestija AN SSSR (1931)  VII, Nr.4, 491-539 (in Russian).


\bibitem{lanczos}
C.\, Lanczos, 
``An iteration method for the solution of the eigenvalue problem of linear differential 
and integral operators," J. Res. Natl. Bur. Stand. {\bf  45} (1950) 255.

\bibitem{os12}
S.\, Odake and R.\, Sasaki,
``Orthogonal Polynomials from Hermitian Matrices,"
J. Math. Phys. {\bf 49} (2008) 053503 (43 pp),
{\tt arXiv:0712.4106[math.CA]}. 

\bibitem{os8}
S.\, Odake and R.\, Sasaki,
``Exact solution in the Heisenberg picture and annihilation-creation operators,"
Phys. Lett. {\bf B641} (2006) 112--117,
{\tt arXiv:quant-ph/0605221}.

\bibitem{os7}
S.\, Odake and R.\, Sasaki,
``Unified Theory of Annihilation-Creation Operators for Solvable (`Discrete') Quantum Mechanics,"
J. Math. Phys. {\bf 47} (2006) 102102, 33pages,
{\tt arXiv:\hspace{0pt}quant-ph/0605215}.


\bibitem{vis}
V.\,S.\, Viswanath and G.\, M\"uller,
 ``The Recursion Method: Applications to Many-body Dynamics," Springer, (2008).
 
 \bibitem{os24}
S.\, Odake and R.\, Sasaki,
``Discrete quantum mechanics," (Topical Review)
J. Phys. {\bf A44} (2011) 353001 (47 pp),
{\tt  arXiv:1104.0473[math-ph]}.


\bibitem{nieto1}
M.\, M.\, Nieto and L.\,M.\, Simmons, Jr.,
``Coherent States For General Potentials,''
Phys. Rev. Lett. {\bf 41} (1978) 207-210.
\bibitem{nieto2}
M.\, M.\, Nieto and L.\,M.\, Simmons, Jr.,
``Coherent States For General Potentials," 1. Formalism,
Phys. Rev. D {\bf 20} (1979) 1321-1331.
\bibitem{nieto3}
M.\, M.\, Nieto and L.\,M.\, Simmons, Jr.,
``Coherent States For General Potentials,"
2. Confining One-Dimensional Examples,
Phys. Rev. D {\bf 20} (1979) 1332-1341.
\bibitem{nieto4}
M.\, M.\, Nieto and L.\,M.\, Simmons, Jr.,
``Coherent States For General Potentials,"
3. Nonconfining One-Dimensional Examples,
Phys. Rev. D {\bf 20} (1979) 1342-1350.




\bibitem{os13}
S.\, Odake and R.\, Sasaki,
``Exactly solvable `discrete' quantum mechanics;
shape invariance, Heisenberg solutions,
annihilation-creation operators and coherent states,"
Prog. Theor. Phys. {\bf 119} (2008) 663-700,
{\tt arXiv:0802.1075[quant-ph]}.



\bibitem{os9}
S.\, Odake and R.\, Sasaki,
``Exact Heisenberg operator solutions for multi-particle quantum
       mechanics,"
J. Math. Phys. {\bf 48} (2007) 082106, (12 pp), 
{\tt  arXiv:0706.0768v1[quant-ph]}.




 \bibitem{mpoly1}
 R.\, Sasaki,
 ``Multivariate Kawtchouk polynomials as  Birth and Death polynomials,"
 {\tt arXiv:2305.08581v2\hspace{0pt}[math.CA]}.
  \bibitem{mpoly2}
 R.\, Sasaki,
 ``Multivariate Meixner polynomials as  Birth and Death polynomials,"
{\tt arXiv:2310.\hspace{0pt}04968[math.CA]}.
 \bibitem{mpoly3}
 R.\, Sasaki, 
 ``Rahman polynomials,"
 {\tt arXiv:2310.\hspace{0pt}17853v2[math.PR]}.
 
\end{thebibliography}
\end{document}